\documentclass[12pt,a4paper]{article}            
 \usepackage[skins,theorems]{tcolorbox}
\tcbset{highlight math style={enhanced,
  colframe=red,colback=white,arc=0pt,boxrule=1pt}}
  \usepackage[bookmarksopen, bookmarksnumbered, bookmarksopenlevel=2]{hyperref}
  \usepackage{tikz}
  \usepackage{tikz-3dplot}
 \usetikzlibrary{calc}
 \usetikzlibrary{decorations} %
 \usepackage[UKenglish]{babel}
 \usepackage[toc,page]{appendix}
 \usepackage{amsmath}
 \usepackage{amssymb}
 \usepackage{graphicx}
 \usepackage{hhline}
 \usepackage[bf]{caption}
\usepackage{cite}
\usepackage[vcentermath]{youngtab}
\usepackage{geometry}
\usepackage{slashed}
\usepackage{color}
\usepackage{stackrel}
\usepackage{tikz-cd} 
\usepackage{mathtools}
\usepackage{cancel} 
\usepackage{multirow}

\usepackage{empheq}
\usepackage{arydshln}

\newcommand{\bqa}{\begin{eqnarray}}
\newcommand{\eqa}{\end{eqnarray}}

\def\C{\mathbb{C}}

\newenvironment{eqn}{\begin{equation}\begin{aligned}}{\end{aligned}\end{equation}\noindent}
\newenvironment{eqn*}{\begin{equation*}\begin{aligned}}{\end{aligned}\end{equation*}\noindent}
\hypersetup{
    pdftitle={},
    pdfauthor={},
    pdfsubject={}
}
\numberwithin{equation}{section}
\numberwithin{table}{section}

\makeatletter

\newcommand{\be}{\begin{equation}}
\newcommand{\ee}{\end{equation}}
\newcommand{\beq}{\begin{equation}}
\newcommand{\eeq}{\end{equation}}
\newcommand{\ba}{\begin{aligned}}
\newcommand{\ea}{\end{aligned}}

\newcommand{\bea}{\begin{eqnarray}}
\newcommand{\eea}{\end{eqnarray}}

\newcommand{\cN}{\mathcal{N}}

\newcommand{\cM}{\mathcal M}

\newcommand\bi{\begin{itemize}}
\newcommand\ei{\end{itemize}}

\renewcommand{\b}{{\beta}}

\def\Im{\mathop{\mathrm{Im}}\nolimits}

\def\Re{\mathop{\mathrm{Re}}\nolimits}

\def\unit{{1\kern-.65ex {\rm l}}}
\def\1{{1\kern-.65ex {\rm l}}}

\newcount\hour \newcount\minute
\hour=\time \divide \hour by 60
\minute=\time
\def\now{%
\ifnum \hour<13
  \ifnum \hour=0 \advance \hour by 12 \number\hour:\else \number\hour:\fi%
     \ifnum \minute<10 0\fi%
     \number\minute%
\ A.M.%
\else \advance \hour by -12 \number\hour:%
  \ifnum \minute<10 0\fi%
  \number\minute%
  \ P.M.%
\fi%
}

\makeatother

\begin{document}

\begin{titlepage}
\begin{center}
\rightline{\small }

\vskip 15 mm

{\large \bf
Holography and the KKLT Scenario
} 
\vskip 11 mm

Severin L\"ust,$^{1}$ Cumrun Vafa,$^{1}$ Max Wiesner,$^{2}$ Kai Xu$^{3}$

\vskip 11 mm
\small ${}^{1}$ 
{\it Jefferson Physical Laboratory, Harvard University, Cambridge, MA 02138, USA}  \\[3 mm]
\small ${}^{2}$ 
{\it Center of Mathematical Sciences and Applications, Harvard University,\\ Cambridge, MA 02138, USA}
 \\[3 mm]
\small ${}^{3}$
{\it Department of Mathematics, Harvard University, Cambridge, MA 02138, USA}

\end{center}
\vskip 17mm

\begin{abstract}
The KKLT scenario, one of the few ideas to realize dS vacua in string theory, consists of two steps: the first involves the construction of a supersymmetric AdS vacuum with a small negative cosmological constant, and the second involves breaking supersymmetry and uplifting the energy to achieve dS.  In this paper we use conventional holography to argue why it is not possible to complete the first step.  We obtain this by putting a bound on the central charge of the dual theory which involves branes wrapping special Lagrangian cycles in CY 4-folds. We find that $l_{\rm AdS}^2 \lesssim{\chi(CY_4)}$.  Since $l_{\rm species}^2\gtrsim \chi(CY_4)$ this leads to $l_{\rm AdS}/l_{\rm species}\lesssim 1$ leading at best to a highly curved AdS which is beyond the validity of the EFT.
\end{abstract}

\vfill
\end{titlepage}

\newpage

\tableofcontents

\setcounter{page}{1}
\section{Introduction}
Our universe seems to realize a positive cosmological constant which, at least to a good approximation, appears not to change in time.  Therefore the late time universe should be described by a quasi-de Sitter geometry. It is thus important to try to construct quasi-de Sitter solutions in string theory.  This turns out to be rather difficult as has been discussed in the context of the Swampland program \cite{Obied:2018sgi,Bedroya:2019snp}.  The main issue is that in the weak coupling limit where we have perturbative control, one typically finds a runaway exponential potential which is expected based on the combination of the distance conjecture \cite{Ooguri:2006in}, which predicts a tower of exponentially light states together with the fact that potentials are related to some power of the mass scale of the tower.  This is also a manifestation of the Dine-Seiberg problem \cite{Dine:1985he}.  Moreover in this limit the slope of the potential $|V'/V|$ is too big \cite{Bedroya:2019snp,Bedroya:2019tba,Rudelius:2021oaz} to be consistent with a quasi-dS solution.  Thus we do not expect to have a dS solution at arbitrarily weak coupling.  If there exists a quasi-dS solution, it should be realized in the interior of the field space, perhaps corresponding to strongly coupled points, for which we currently have rather limited analytic tools.  Nevertheless, there has been claims in the literature  (see e.g. \cite{Kachru:2003aw,Balasubramanian:2005zx}) that quasi-dS solutions are realizable in the weak (but not arbitrarily weak) coupling points.   This is not ruled out by the asymptotic values of the potential and it would be an interesting possibility if one can indeed realize such vacua in string theory.  In this paper we focus our attention on what is viewed as one of the most promising such attempts, namely the KKLT scenario and argue why it is not possible to realize this scenario.

The KKLT scenario involves studying flux vacua of Type IIB on orientifolds of Calabi-Yau 3-folds (or, more generally, F-theory on elliptic 4-folds).  The existence of fluxes and 3-branes is generally forced on us by the tadpole condition on the 3-brane charge which receives a contribution from the CY 4-fold given by $-\chi/24$, where $\chi$ is the Euler characteristic of the CY 4-fold \cite{Becker:1996gj,Sethi:1996es}.  The KKLT scenario attempts to use the large number of possible fluxes to find a vacuum which is supersymmetric and realizes a small negative value for the cosmological constant. It then uses spacetime filling anti-D3 branes at highly warped points of the geometry to lift the solution to dS.
To achieve the second step we need to start with a sufficiently small negative cosmological constant of AdS in the first step, as the uplift is of the same order as the absolute value of the cosmological constant of AdS.
A lot of the focus on the KKLT scenario has been on checking the validity of the second step even though some issues regarding the first step have also been raised recently, questioning e.g. the possibility to find flux vacua within the tadpole bound for large number of moduli \cite{Bena:2020xrh} or raising the issue that strong warping yields to large singular regions of the CY orientifold \cite{Gao:2020xqh}. Other issues have also been raised (see e.g. \cite{Sethi:2017phn}). The consistency of the second step is more difficult to establish in particular due to the breaking of the supersymmetry.  Here we focus on the first step, where a supersymmetric solution with small negative cosmological constant is desired.
We show that this is not possible and estimate that the AdS length scale in Planck units is bounded as $l_{\rm AdS}^2<\chi$ where $\chi$ is the Euler characteristic of the CY 4-fold.
Moreover since $\chi$ is a measure of the number of light degrees of freedom coming from the CY compactification, we learn that $l_{\rm species}^2\gtrsim \chi$ \cite{Dvali:2007hz,Dvali:2007wp,Dvali:2010vm}.  In other words even for large $\chi$, we find $l_{\rm AdS}\lesssim l_{\rm species}$.  This shows that the EFT breaks down and therefore we cannot trust the solution.%

The argument we employ to show this is holography.  Since in the first step one would like to achieve a small cosmological constant, we should use the condition of having large values for the fluxes to give us a large statistical possibility \cite{Bousso:2000xa,Douglas:2003um,Denef:2004cf}.  Therefore this should lead to a holographic realization in the usual way.  In particular, since we are in the large flux regime, we would dualize this to the corresponding branes which are 2+1 branes consisting of bound states of D5/NS5-branes wrapped around 3-cycles of the CY, as has been previously suggested in \cite{Silverstein:2003jp}.\footnote{Properties of the CFT dual to the KKLT AdS vacuum have also been discussed in \cite{deAlwis:2014wia}.}  In the M-theory context these map to M5 branes wrapping special Lagrangian (SLag) submanifolds.  We argue why the degrees of freedom on these branes do not grow any faster than $\chi$ of the CY, thus leading to our bound.

The organization of this paper is as follows:  In section 2 we review the KKLT scenario.  In section 3 we describe the holographic dual theory.  In section 4 we study SLags and their deformations to put a bound on the central charge of the dual theory. In section 5 we show how these results add up to an obstruction for the KKLT scenario.  Moreover in that section we explain in detail why some of the recent attempts to realize the first step of KKLT \cite{Demirtas:2019sip,Demirtas:2020ffz,Blumenhagen:2020ire,Demirtas:2021nlu,Demirtas:2021ote} will not lead to supersymmetric AdS vacua as was hoped for.
In section 6 we present our conclusions.

\section{Review of the KKLT scenario}\label{sec:KKLT}

In this section we want to briefly review the key aspects of the KKLT scenario \cite{Kachru:2003aw} pertinent to the analysis of this paper. This scenario proposes that a dS vacuum can be obtained from type IIB/F-theory flux compactifications by applying a series of steps. Therefore consider F-theory compactified on an elliptically fibered Calabi--Yau fourfold $X_4$ and further take the orientifold limit where $X_4$ is given by $X_4 = (X_3 \times T^2)/\mathbb{Z}_2$. Here, $X_3$ is a Calabi--Yau threefold and the $\mathbb{Z}_2$ acts as $(-1)^{F_L} \Omega_p \sigma$ where $F_L$ is the fermion number in the left-moving sector, $\Omega_p$ worldsheet parity and $\sigma$ a holomorphic involution on $X_3$. The latter acts on the holomorphic three-form as $\sigma^* \Omega_3=-\Omega_3$. The resulting fourfold $X_4$ is singular since in general $\sigma$ has fixed loci which can be interpreted as O3 and O7 planes. The presence of the orientifold planes induces a D3-brane tadpole that needs to be cancelled either by space-time filling three-branes or by a combination of the type IIB RR and NS threeform fluxes $F_3$ and $H_3$. 

Describing the F-theory compactification on $X_4=(X_3\times T^2)/\mathbb{Z}_2$ times a circle via the dual M-theory on $X_4$, the type IIB three-form fluxes $F_3$ and $H_3$ map to different components of the M-theory four-form flux $G_4$ 
\begin{align}\label{G4flux}
    G_4 = F_3 \wedge a + H_3 \wedge b \,, 
\end{align}
where $a,b$ are the 1-forms on the $T^2$ dual to the $A$- and $B$-cycle of $T^2$ for which we have 
\begin{align}
    1 = \int_{A} \Omega_1 \,,\qquad \tau = \int_{B} \Omega_1 \,.
\end{align}
Here $\Omega_1$ is the holomorphic $(1,0)$ form on $T^2$ and $\tau$ its complex structure parameter. The $G_4$ flux needs to satisfy the quantization condition \cite{Witten:1996md}
\begin{align}
    G_4 +\frac{c_2(X_4)}{2} \in H^4(X_4,\mathbb{Z})\,. 
\end{align}
In the three dimensional effective theory obtained by compactifying M-theory on $X_4$ the M2-brane tadpole cancellation condition reads \begin{align}\label{eq:Mtheorytadpole}
       \frac{\chi(X_4)}{24}= N_{\rm M2} +\frac{1}{2}\int_{X_4} G_4\wedge G_4\,,  
\end{align}
with $\chi(X_4)$ the Euler characteristic of the fourfold $X_4$ and $N_{\rm M2}$ the number of space-time filling M2-branes. Lifting the 3d M-theory to four-dimensional type IIB/F-theory, the M2-branes get mapped to $N_{\rm D3}=N_{\rm M2}$ space-time filling D3-branes and using \eqref{G4flux}, the tadpole cancellation now reads 
\begin{align}\label{IIBtadpole}
       \frac{\chi(X_4)}{24}= N_{\rm D3} +\frac12 \int_{X_3} F_3\wedge H_3\,,  
\end{align}
The $G_4$-flux induces the superpotential \cite{Gukov:1999ya} 
\begin{align}\label{GVWG4}
    W_{\rm GVW} = \int_{X_4} G_4\wedge \Omega_4\,,
\end{align}
where $\Omega_4$ is the holomorphic $(4,0)$ form on $X_4$. In the case $X_4 = (X_3 \times T^2)/\mathbb{Z}_2$ we have $\Omega_4 = \Omega_3\wedge \Omega_1$\,, such that in 4d the superpotential reads 
\begin{align}\label{GVW}
    W_{\rm GVW} = \int_{X_3} G_3\wedge \Omega_3 \,, \qquad G_3 =F_3- \tau H_3\,. 
\end{align}
A general superpotential $W$ in a $\cN=1$ theory of supergravity induces a scalar potential given by 
\begin{align}\label{Vfull}
    V = e^K \left(g^{a\bar b} D_a W \bar{D}_{\bar b} \bar W -3 |W|^2 \right)\,,
\end{align}
where $K$ is the K\"ahler potential, $g_{a\bar b}=\partial_a \partial_{\bar b} K$ the metric on the moduli space and $D_a = \partial_a + \partial_a K$ the K\"ahler covariant derivative. Here, $a,b$ run over all scalar fields of the effective field theory including both, the complex structure deformations $z^i$, $i=1,\dots h^{2,1}_-$ and the (complexified) K\"ahler deformations $T_\alpha$, $\alpha=1,\dots h^{1,1}_+$, as well as the axio-dilaton $\tau$. At tree-level (weak string coupling) the 4d K\"ahler potential in the large volume limit is given by 
\begin{align}
    K = -\log \left(-i (\tau-\bar\tau)\right) - \log\left(-i \int_{X_3} \Omega_3 \wedge \bar \Omega_3 \right) -2 \log(\mathcal{V})\,,
\end{align}
where $\mathcal{V}$ is the classical volume of $X_3$. The GVW superpotential \eqref{GVW} does not depend on the K\"ahler deformations such that, using the no-scale property $K_{\alpha} K_{\bar \beta} g^{\alpha \bar \beta} =3$\,, the scalar potential simplifies to
\begin{align}
    V = e^K g^{I \bar J} D_I W \bar{D}_{\bar J} \bar W\,,
\end{align}
where here $I,J$ run over complex structure deformations and the axio-dilaton. The minima of this potential correspond to the solutions of the F-term equations $D_I W=0$, $\forall I$ and thus necessarily have $V=0$. Given the decomposition of the $H^4(X_4,\mathbb{C})$ as 
\begin{align}
    H^4 = H^{4,0}\oplus H^{3,1}\oplus H^{2,2}\oplus H^{1,3} \oplus H^{0,4}\,,
\end{align}
the F-term condition $D_I W=0$ for the superpotential in \eqref{GVWG4} implies that $G_4$ has to be self-dual. Therefore $G_4$ has to satisfy
\begin{align}
    G_4 \in H^{4,0}\oplus  H^{2,2}_+\oplus  H^{0,4}\,, 
\end{align}
where $H^{2,2}_+$ is the self-dual part of $H^{2,2}$. For a given quantized flux $G_4$ this condition constrains the possible choice of complex structure and therefore renders some complex structure deformations massive. For a supersymmetric vacuum, we further need to require $W=0$ which implies $G_4\in H^{2,2}$ and primitivity, i.e.\ $G_4\wedge J=0$ with $J$ the K\"ahler form on $X_4$. In terms of the type IIB flux $G_3$ the above conditions of $G_4$ translates into $G_3$ being imaginary self-dual and of $(2,1)$ type whereas for non-supersymmetric solutions $G_3$ also can have a $(0,3)$ component. 

The starting point of the KKLT scenario is a flux configuration such that at the point in complex structure moduli space at which $D_I W=0$ the $G_3$ flux has a non-vanishing $(0,3)$ component (or the $G_4$ flux has a $(0,4)$ component) such that $W_0 \equiv W|_{D_I W=0}\neq 0$. As a consequence the F-term equation for the K\"ahler moduli $D_{T_\alpha} W \propto W$ being zero is not satisfied. It is further assumed that the locus in complex structure moduli space where $G_3$ is imaginary self-dual ($G_4$ is self-dual) consists of isolated points such that all complex structure deformations and the dilaton are massive. However, the K\"ahler directions remain as flat directions and need to be stabilised by taking into account non-perturbative corrections to the superpotential from D3-brane instantons wrapped on divisors of $X_3$. These non-perturbative corrections have the form 
\begin{align}\label{eq:Wnp}
    W_{\rm n.p.} = \sum_{\mathbf{k}} \mathcal{A}_{\mathbf{k}}(z^i, G)\, e^{-2\pi k^\alpha T_\alpha }\,.
\end{align}
Here $\mathbf{k}$ scans through effective divisors of $X_3$ and the Pfaffian determinant $\mathcal{A}$ depends in general on the complex structure deformations $z^i$ (including $\tau$) and the choice of flux $G$. After integrating out the massive complex structure deformations, following the KKLT scenario, $\mathcal{A}$ can effectively be treated as a constant (assuming the complex structure dependence of $W_{\rm n.p.}$ is mild enough) such that the full superpotential reads 
\begin{align}\label{spotwithW0}
    W = W_0 +  \sum_{\mathbf{k}} \mathcal{A}_{\mathbf{k}}^{\rm eff}\, e^{-2\pi k^\alpha T_\alpha }\,. 
\end{align}
Since the superpotential now depends on the K\"ahler moduli, it is possible to also solve the F-term equations for the K\"ahler directions. Following the original discussion \cite{Kachru:2003aw}, let us take the simplifying assumption that there is a single K\"ahler modulus $T$ and consider a single non-perturbative contribution to the superpotential $W_{\rm n.p.}= \mathcal{A} \,e^{-2\pi a T}$. We further set $\Im T=0$ and $\Re T=\sigma$. In this case the F-term equation $D_T W=0$ implies 
\begin{align}
    W_0 = - \mathcal{A} \,e^{-a \sigma_0}\left(1+\frac{2}{3}a \sigma_0\right)\,,
\end{align}
where $\sigma_0$ is the value of $\sigma$ at the critical point. Thus, in order for the instanton expansion and $\alpha'$-corrections to be under control, i.e.\ $\sigma_0\gg 1$, one requires $W_0$ to be exponentially small.  Given the enormity of possible directions it can be hoped that this can be indeed achieved \cite{Ashok:2003gk}. By \eqref{Vfull} the value of the potential (in Planck units) at the critical point is given by 
\begin{align}
    V_0 = \left.-3 \left(e^K |W|^2\right)\right|_{D_aW=0} = -\frac{a^2 \mathcal{A}^2 e^{-2a \sigma_0}}{6 \sigma_0} <0\,.
\end{align}
Assuming all this works as planned, one expects a supersymmetric AdS vacuum with exponentially small cosmological constant
\begin{align}
    \Lambda = V_0 M_{\rm pl}^2 \ll 1\,,
\end{align}
since $\sigma_0 \gg 1$ by assumption. Starting from this supersymmetric AdS vacuum with a very small cosmological constant, in a next step the KKLT scenario then proposes an uplift to dS by means of $\overline{D3}$ branes. For the analysis of this paper, we are only interested in the first step of the KKLT scenario in which a supersymmetric AdS vacuum with exponentially small cosmological constant is obtained.\footnote{The literature sometimes refers to the complex structure moduli stabilisation as the first step of the KKLT scenario and to the K\"ahler moduli stabilisation by means of non-perturbative corrections as the second step. Throughout this work we refer to the combination of both as the first step of the KKLT scenario. For us the second step would be the uplift of the SUSY AdS vacuum to a dS vacuum.} Therefore, we are not discussing the uplift step of the KKLT scenario here.   Note that regardless of the argument, in order to proceed with the second step of the KKLT scenario we need $|\Lambda |\ll 1$ (as the uplift energy is small as it comes from anti-D3 branes in the warped throat) to lead to dS vacua.

Notice that the assumption of this scenario that the criticality of the superpotential with respect to complex and K\"ahler moduli deformations are decoupled is hard to justify.  Therefore whether or not the scenario can be carried out as hoped is unclear.  Indeed as we will argue in this paper, even if an AdS supersymmetric vacuum is constructed along these lines, one cannot achieve $|\Lambda| \ll 1$.

\section{Holographic dual}\label{sec:holography}
\begin{figure}
\centering
\includegraphics[width=.75\textwidth]{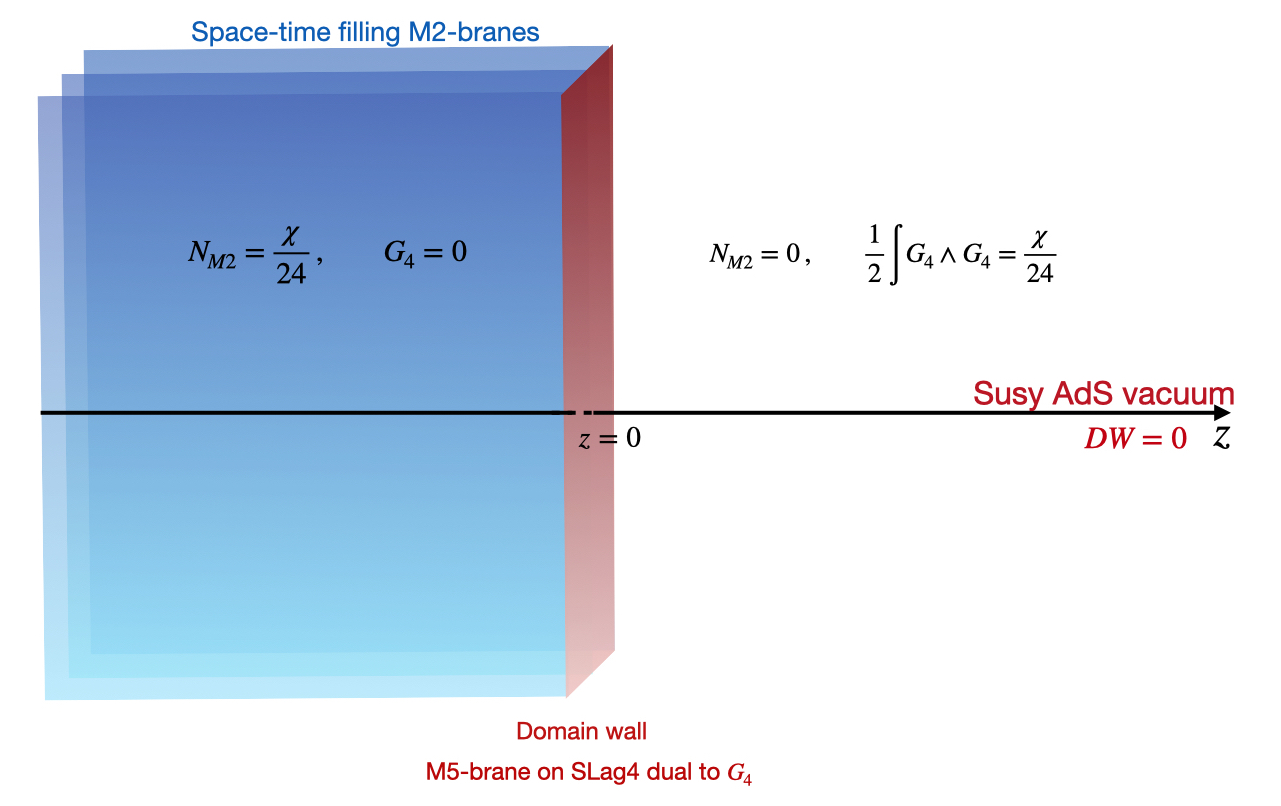}
\caption{The domain wall setup corresponding to an M5-brane wrapped on a Special Lagrangian 4-cycle dual to $G_4$ located at $z=0$ and interpolating between a vacuum with $G_4=0$ and  $\frac{\chi}{24}$ space-time filling M2-branes at $z\rightarrow -\infty$ and the supersymmetric AdS flux vacuum ($DW=0$) for $G_4\neq 0$ at $z\rightarrow \infty$. For the case of supersymmetric AdS$_4$ vacua in type IIB orientifold flux compactifications the picture is similar with M2-branes replaced by D3-branes and the domain wall being made of D5/NS5 branes.}
\label{fig:domainwall}
\end{figure}

In this section, we want to discuss the holographic dual of the supersymmetric AdS vacua obtained from flux-compactifications.  Indeed since one is looking for $|\Lambda| \ll 1$ in AdS$_4$ the theory would be expected to have a microscopic 2+1 dimensional dual. Instead of directly discussing the dual of the AdS$_4$ vacua, let us first consider the 3d version of KKLT, which starts with the case of M-theory compactified on a Calabi--Yau fourfold $X_4$ down to three dimensions, in the presence of a quantized $G_4$ flux. In this case, one aims to find supersymmetric AdS$_3$ vacua with $|\Lambda|\ll 1$ dual to a microscopic 1+1 dimensional theory. For now we assume that indeed the $G_4$ flux allows for a three-dimensional supersymmetric AdS vacuum with radius, as measured in 3d Planck units,
\begin{align}
    \frac{1}{l_{{\rm AdS}_3}^2} = -4  e^K |W|^2 \Bigr|_{D_a W=0}\,,
\end{align}
where $K$ is the full K\"ahler potential including the complex structure and the K\"ahler sector and $W$ is the superpotential including all non-perturbative corrections (arising from Euclidean M5-branes)\footnote{In principle the 3d effective theory also allows for a second kind of superpotential \cite{Gukov:1999ya} $\tilde{W} = \int_{X_4} J_{4} \wedge J_{4} \wedge G_4$. Since we are interested in supersymmetric vacua that have an F-theory lift to four-dimensions, here we are only interested in $G_4$ fluxes that are primitive, i.e. $J_{4}\wedge G_4=0$ for which $\tilde{W}$ vanishes identically.}
\begin{align}\label{fullsuperpot}
     W = \int_{X_4} \Omega_{4} \wedge G_{4} + \sum_{\mathbf{k}} \mathcal{A}_{\mathbf{k}}(z^i, G_4)\,\, e^{-2\pi k^\alpha T_\alpha }\,. 
\end{align}
Associated to the flux we can consider a domain wall in three dimensions obtained by wrapping an $M5$-brane on the Poincar\'e dual four-cycle $L\in H_4(X_4)$ of $G_4$, which we assume saturates the tadpole condition (to obtain the smallest possible $|\Lambda|$). Let us locate the domain wall at $z=0$ with $z$ being the coordinate transverse to it. Then the domain wall dual to the flux $G_4$ interpolates between a vacuum with vanishing flux quanta at $z=-\infty$, and the supersymmetric AdS flux vacuum for $z=\infty$. Notice that in the half-space $z<0$ this configuration requires $\chi(X_4)/24$ space-time filling M2-branes to satisfy the tadpole cancellation condition \eqref{eq:Mtheorytadpole}. We illustrated the domain wall setup in figure~\ref{fig:domainwall}. Alternatively, we can consider the situation where the domain wall caps off the space to the right corresponding to a bubble of nothing on the left. In this case, we do not need any M2-branes on the left. This is essentially the holographic picture where the brane is at the boundary of AdS.

Since we want the AdS vacuum to be supersymmetric, we need the corresponding domain wall to be $\frac{1}{2}$-BPS. In our present case, this means that the 4-cycle $L$ wrapped by the M5-brane has to be special Lagrangian or a holomorphic 4-cycle.  For the uplift to 4d the case of interest would be the special Lagrangian case as we will now argue (the holomorphic 4-cycle case will also be discussed later in this paper for the AdS$_3$ case).
A real four-dimensional submanifold $L$ of $X_4$ is called special Lagrangian if it satisfies the conditions
\begin{equation}\begin{aligned}\label{eq:slag}
J_4 \bigl|_L &= 0 \,, \\
\Im \left(e^{i\alpha} \Omega_4 \right) \bigl|_L &= 0  \,,
\end{aligned}\end{equation}
for some constant phase $\alpha$, 
with $J_4$ the K\"ahler form on $X_4$ and $|_L$ denotes the pull back on $L$.
Accordingly, the world-volume theory on the domain wall preserves $\cN=(1,1)$ supersymmetry in two dimensions. Let us consider the backreaction of this $\frac{1}{2}$-BPS domain wall, following \cite{Cvetic:1992bf,Cvetic:1992st,Ceresole:2006iq,Bandos:2018gjp,Lanza:2020qmt}. For the metric in the three extended space-time directions one makes the Ansatz
\begin{align}
    ds^2 = e^{2D(z)}(-dt^2 +dx^2) +dz^2\,. 
\end{align}
The BPS equations can now be rephrased in terms of the flow equations 
\begin{eqn}\label{eq:BPSflow}
    \frac{d D}{d z} &= -\zeta |\mathcal{Z}| \,,\\
    \frac{d \phi^a}{d z} &=  2 \zeta  g^{a \bar b} \partial_{\bar b}|\mathcal{Z}| \,,
\end{eqn}
where $\zeta=\pm 1$ determines which of half of the supersymmetries is preserved by the flow. Furthermore $\phi^a$ stands for the scalar fields in the theory and we have defined 
\begin{align}\label{eq:Z}
    |\mathcal{Z}| = e^{K/2} |W|\,,
\end{align}
which can be identified as the tension of the domain wall in 3d Planck units. We can interpret the flow equations as up or down gradient flows for  $|\mathcal{Z}|$.   In order for the flow to lead to an AdS vacuum for $z\rightarrow \infty$, $\mathcal{Z}$ needs to be asymptotically constant, $\partial_a |\mathcal{Z}|=0$, and non-zero which implies 
\begin{align}
    D_a W=0 \,,
\end{align}
i.e.\ the F-term condition. Notice that here $W$ is the full superpotential appearing in \eqref{fullsuperpot} including the non-perturbative instanton corrections. Accordingly, the index $a$ also scans over all scalar fields including the K\"ahler moduli. From the domain wall perspective the non-perturbative contributions to $W$ should be interpreted as corrections to the tension of the BPS domain wall which is therefore also sensitive to the K\"ahler moduli. However, under the assumption that the KKLT scenario works, the non-perturbative terms in \eqref{fullsuperpot} are small at the attractor point. Therefore the attractor flow in the complex structure sector should effectively be independent of the non-perturbative corrections. Hence in particular the calibration condition for the BPS domain wall should indeed correspond to the special Lagrangian condition introduced above.

We can thus describe the flux vacua as an attractor flow \cite{Kallosh:2005ax,Ceresole:2006iq} for the domain wall obtained by wrapping an M5-brane on the special Lagrangian cycle $L$ dual to the $G_4$-flux, where the flow is driven by the gradient flow of its tension.\footnote{The relation between supersymmetric AdS$_4$ vacua and the attractor flow for BPS domain walls obtained from branes on calibrated cycles has previously been noticed in \cite{Kounnas:2007dd}.}  The fact that the dual $G_4$ is primitive follows from the condition that $L$ is a Lagrangian submanifold. 
Note that minima and maxima of $|\mathcal{Z}|$ are stable attractor points while the saddle points, even though formally preserving supersymmetry, are not stable attractor values.\footnote{It is tempting to speculate that saddle points of $|\mathcal{Z}|$ cannot occur in a consistent $N=1$ supergravity, and that only the minima of $|\mathcal{Z}|$ can arise, as a Swampland constraint.} Notice that the central charge of the domain wall tension is extremized far away from its location. This is to be contrasted with the attractor mechanism for black holes in $d>3$ \cite{Ferrara:1995ih} where the attractor value for the central charge is reached in the near-horizon regime. This difference arises since domain walls are co-dimension one objects and therefore can be associated to relevant operators whereas black holes in $d>3$ are associated to irrelevant operators, cf. \cite{Lanza:2020qmt} for a more detailed discussion. Note also that in the situation we are concerned with here, where the $G_4$-flux jumps when crossing the domain wall, the flow \eqref{eq:BPSflow} interpolates between the vacua of two different flux-potentials, as explained in detail in \cite{Bandos:2018gjp}.

From the above discussion we infer that the supersymmetric AdS$_3$ vacuum obtained from M-theory on $X_4$ in the presence of $G_4$-flux is holographically dual to a CFT realized on the world-volume of the domain wall obtained by wrapping an M5-brane on the special Lagrangian four-cyle $L$ dual to the $G_4$-flux. More precisely, interpreting the coordinate $z$ transverse to the domain wall as holographic RG-scale, the CFT dual to the supersymmetric AdS$_3$ vacuum is realized as the IR limit of the world-volume theory of the M5-brane on $L$. Through the AdS$_3$/CFT$_2$ dictionary we can now relate the AdS$_3$ radius, $l_{\rm AdS}$, to the central charge of the IR theory, $c_{\rm IR}$, as 
\begin{align}\label{Lambdac}
    c_{\rm IR} =\frac{3}{2}\, l_{\rm AdS}\,.
\end{align}
We can thus identify the attractor value of the domain wall tension with the IR value of the central charge of the CFT on its world-volume 
\begin{align}
    |\mathcal{Z}_{\rm AdS_3}|^2 = \frac{9}{16 c_{\rm IR}^2} \,. 
\end{align}
In order to find the cosmological constant of the AdS$_3$ vacuum we thus need to count the IR degrees of freedom of the world-volume theory of the M5-brane on $L$ since $\Lambda_3\sim {-1/c_{IR}^2}$. However, instead of counting the IR degrees of freedom, in the following we will rather count the UV degrees of freedom on the M5-brane wrapped on $L$ which can be inferred from the geometry of $L$. The $c$-theorem tells us that under RG-flow the degrees of freedom of the 2d field theory on the world-volume cannot increase such that 
\begin{align}
    c_{\rm UV} \geq c_{\rm IR}\,,
\end{align}
which we can use to give a lower bound on the cosmological constant via \eqref{Lambdac}.  The UV degrees of freedom of the worldvolume theory of the M5-brane on $L$ will be discussed in the next section.  As we will see there this counting is purely topological. Therefore it does not depend on the precise point in moduli space. However, when going to the IR some degrees of freedom will be lifted as a consequence of the non-perturbative corrections to $W$ and therefore $c_{\rm IR}$ indeed is sensitive to the point in moduli space. In particular only at the attractor point corresponding to $\partial_a |\mathcal{Z}|=0$ does the IR worldvolume theory on $L$ correspond to a CFT. Similarly in the 4d case we would be studying the degrees of freedom on D5/NS5 brane domain walls and the 4d AdS scale would be expected to scale as the number of degrees of freedom $c$ on the domain wall, leading to $\Lambda_4 \sim -1/c_{\rm IR}$.\\

Let us note that already in \cite{Silverstein:2003jp} the holographic dual of AdS$_4$ vacua in type IIB flux compactifications was described by trading the flux for D5-/NS5-branes wrapping the dual cycles. Again, the worldvolume theory on the branes is identified with the CFT dual of the AdS$_4$ vacuum. However, \cite{Silverstein:2003jp} employs a description where the 5-branes are separated in the radial AdS direction with D3-branes stretched between them to account for the D3-brane charge. This setup describes the Coulomb branch of the CFT where some degrees of freedom become massive. In \cite{Silverstein:2003jp} these degrees of freedom are counted by string junctions ending on the different sets of 5- and 3-branes. In this paper, however, we take a different approach by treating the different stacks of 5-branes as a single object in M-theory, i.e.\ the M5-brane wrapped on a special Lagrangian four-cycle. We can therefore work directly at the origin of the Coulomb branch where all CFT degrees of freedom are massless.

\section{Special Lagrangian cycles and their deformations}\label{sec:SLag} 

As we have seen in the previous section the holographic dual to a supersymmetric $\text{AdS}_d\times CY_4$ background is expected to be given by suitable branes wrapped around special Lagrangian (SLag) cycles.  The central charges $c$ of the resulting CFTs are determined by the deformation of these cycles.  The central charges for dimensions $d>2$ scale as $c\propto l_{\rm AdS}^{d-2}$, where $l_{\rm AdS}$ is the AdS radius in Planck units.\footnote{For the $d=2$ case the central charge can be viewed as the entropy of the black hole, i.e.\ the volume of the horizon $c\propto {\rm vol} (S^n)\sim {(l_{\rm AdS})^n}$.} Thus any bound on the deformations of the dual special Lagrangians restricts the value of the corresponding cosmological constant which scales as $\Lambda =-1/l_{\rm AdS}^2\sim {- c^{\frac{-2}{d-2}}}$.   We will be interested in the case of large cycle classes (leading to small cosmological constant) and we would like to estimate how $c$ grows with the cycle class.  In particular, since the number of deformations of the SLag in a Calabi-Yau manifold is given by its first Betti number $b_1$ \cite{Mclean1998DeformationsOC} we would like to estimate how $b_1$ grows upon rescaling the homology class the branes wrap on.

In addition to special Lagrangian cycles it will also be useful to review some of the examples of branes wrapping holomorphic cycles of CY $n$-folds that have arisen in the context of holography. In particular we will briefly review holography of branes wrapping cycles in $CY_2=K3$, $CY_3$- and $CY_4$-folds. The examples we consider are obtained from Type IIB/F-theory as well as M-theory. For type IIB/F-theory we consider 

\vspace{.2em}
\begin{center}
\begin{tabular}{c|c|c|c}
Case & Geometry & Brane & dual field theory\\ \hline 
I & AdS$_3\times S^3\times K3$ & D3-brane on SLag2 or hol$(C_2)$& $d=2$, $\cN=(4,4)$\\ \hline
II& AdS$_2\times S^2\times CY_3$& D3-brane on SLag3 & $d=1$, $\cN=4$\\\hline 
\multirow{2}{*}{III} & AdS$_4\times (X_3\times T^2)/\mathbb{Z}_2$&D5/NS5-branes on SLag3 & \multirow{2}{*}{$d=3$, $\cN = 1$}\\ 
& AdS$_4\times \text{Ell}(CY_4)$ & $[p,q]$ 5-branes in base& \\
\end{tabular}
\end{center}
\vspace{.1em}

\noindent where cases I and II arise in type IIB and for the case III we consider type IIB on CY$_3$ orientifolds or, more generally, F-theory on elliptic $CY_4$ folds. In the case of M-theory we consider:
\vspace{.2em}
\begin{center}
\begin{tabular}{c|c|c|c}
Case & Geometry & Brane  & dual field theory\\ \hline 
IV & AdS$_3\times S^2\times CY_3$ & M5-brane on hol$(C_4)$ & $d=2$, $\cN = (0,4)$ \\ \hline
V$_1$ & \multirow{2}{*}{AdS$_3\times CY_4$}& M5-brane on hol($C_4$) & $d=2$, $\cN = (0,2)$ \\
V$_2$ &  &M5-brane on SLag4 & $d=2$, $\cN = (1,1)$ \\ 
\end{tabular}
\end{center}
\vspace{.1em}
Here, case IV corresponds to M5 branes wrapping holomorphic 4-cycles of $CY_3$ leading to MSW strings \cite{Maldacena:1997de}. For the case V we differentiate two possibilities, V$_{1,2}$: case V$_1$ corresponds to M5-branes wrapping holomorphic 4-cycles leading to $(0,2)$ supersymmetric 2d CFT's whereas in case V$_2$ M5-branes are wrapped on SLag 4-cycles leading to $(1,1)$ supersymmetric 2d CFT's. Before discussing all cases separately, let us briefly summarize some general properties about SLag cycles. 

\subsection{Some generalities about SLags}
A special Lagrangian submanifold (SLag) $L_n$ is a mid-dimensional cycle in a $CY_n$ manifold $X_n$ which is a Lagrangian subspace with respect to the K\"ahler form and for which the restriction of the holomorphic $n$-form $\Omega_n|_{L_n}$  is proportional to the volume form up to an overall phase \cite{Harvey:1982xk}. 
It hence satisfies conditions analogous to \eqref{eq:slag}.
Using this one can show that it minimizes the volume in the corresponding cohomology class and that it leads to $\frac12$-BPS states when supersymmetric branes wrap it. Typically, when we wrap a large number of branes in such a class, or when the class is large even if it is primitive, we expect the attractor mechanism to set in and to change the modulus so that at the attractor value of the complex structure it minimizes the mass/tension of the brane.  As far as the complex structure is concerned this is achieved by extremizing
\begin{align}\label{ZL}
    \left|Z(L_n)\right|^2 =  \frac{\left|\int_{L_n} \Omega_n\right|^2 }{\int_{X_n}   |\Omega_n|^2} = \left|\int_{L_n} \widehat{\Omega}_n \right|^2\,,
\end{align}
where we have introduced the normalized holomorphic $(n,0)$-form
\begin{align}
    \widehat{\Omega}_n = \left(\int_{X_n}\Omega_n\wedge\bar{\Omega}_n\right)^{-\frac12} \Omega_n \,.
\end{align}

For the attractor value of the K\"ahler moduli we expect that, if the mass/tension depends on it, it should settle to the minimum value (in Einstein frame).  This is the case for example for type IIB on $K3$ or $CY_3$.  Moreover in these cases it is easy to check that the tension/mass of D3 branes wrapping SLags does not depend on the overall volume of the CY and thus the attractor mechanism does not fix that.

The attractor value of \eqref{ZL} may or may not be zero even if the wrapped cycle is not movable.  For example for the case of $K3$ if we wrap a genus 0 surface the attractor value of the tension is zero, corresponding to the point in moduli space where the sphere shrinks to a point (the $A_1$ singularity). Similarly if in a $CY_3$ we wrap a conifold class the tension is also minimized at the point in moduli space where this class shrinks to zero size.  In such cases we do not obtain a holographic dual theory. Therefore having a SLag is not enough to guarantee a holographically dual AdS but we also need the attractor value of \eqref{ZL} to be non-zero, $|Z(L)|\not=0$. In particular, a critical point of  $|Z(L)|$ with $|Z(L)| =0$ does not yield an AdS solution. 
 
Notice that for the cases III and V$_2$, i.e.\ the case of F-theory on elliptic $CY_4$ or M-theory on $CY_4$, the tension of a brane wrapped on a SLag4-cycle {\it does depend} on the overall volume and is minimized in the limit $\mathcal{V}\rightarrow \infty$.  So in these cases, if there are no further corrections to the BPS tension, we expect the attractor value to correspond to a decompactification of the $CY_4$.  This is related to the first step of the KKLT scenario, i.e.,  obtaining a supersymmetric AdS vacuum. In this case if there were no corrections to the superpotential depending on K\"ahler moduli, we would not be getting an AdS vacuum at finite volume for the CY.\footnote{In the non-geometric setup of \cite{Becker:2006ks} where there are no K\"ahler moduli this issue does not arise and one can obtain reliable supersymmetric AdS vacua.  However the value of the cosmological constant is close to the Planck value. This is also the case for KKLT scenario as we will argue in this paper.}  It is often said that in such a case without the K\"ahler corrections, there would be no supersymmetric vacuum because in this case $DW\not=0$ in the overall volume direction.  However, this is not entirely correct:  Indeed $DW$ becomes proportional to $W$ which vanishes if and only if we go to $\mathcal{V}\rightarrow \infty$ leading to a non-compact supersymmetric model.  We believe that this is what will happen generically.  However, as reviewed in section \ref{sec:KKLT}, in general there are corrections to the superpotential depending on the K\"ahler moduli and this may affect this statement such that, in principle, one could find a supersymmetric AdS vacuum due to such corrections. If we assumed that the attractor value of the tension is unique, as is the case for all the known supersymmetric attractor cases (for an example see \cite{Wijnholt:1999vk}), this would imply that the minimum is  again at infinite volume (where K\"ahler corrections vanish) with $W=0$, and thus even then there can never be a supersymmetric AdS vacuum which minimizes the domain wall tension. However for the sake of completeness we will assume that there can be multiple attractor values corresponding possibly to saddle points of the tension and aim to find a bound on how small the resulting cosmological constant can be.

\subsection{\texorpdfstring{Case I: SLag2 and hol(C$_2$) in K3}{Case I: SLag2 and hol(C2) in K3}}\label{ssec:n2}

Let us start with case I and consider type IIB on K3, and wrap a D3-brane on a Riemann surface. This leads to a string in 6d.  This is a well known example \cite{Kim:2005ez} of holography.  In this case, since K3 is hyperK\"ahler, SLag2 and holomorphic two-cycles hol($C_2$) are equivalent, depending on the choice of the complex structure of K3.  Let $(P_L,P_R)$ denote the integral even self-dual lattice of the 2-cycles of K3. For a given two-cycle the $R(L)$ components are obtained by projecting to self-dual (anti-self-dual) parts, which depends on the metric. For a Riemann surface $C_2$ in the class $(P_L,P_R)$, which realizes a SLag2 or holomorphic two-cycle, the genus is given by
$$P_R^2-P_L^2=2g-2=(C_2)^2$$
where $(C_2)^2$ is the self-intersection of $C_2$ in $K3$.
The tension of a D3-brane wrapped on $C_2$ in the Einstein frame is given by $P_R=\sqrt{2g-2+P_L^2}$.
For $g=0,1$ the attractor value for the tension can be zero for suitable choices of $P_L$. For $g\geq 2$ the attractor value is non-zero and corresponds to the minimal value of $P_R^2$ which is achieved for $P_L=0$.  In this case the tension $T$ of the D3-brane at the attractor value is given by 
$$T^2= 2g-2=C_2^2 \,. 
$$ 
On the other hand, the number of deformations of the genus $g$ cycle $C_2$ is given by 
\begin{align}
    b_1(C_2)=2g=C_2^2+2\,. 
\end{align}
Therefore, the number of deformations which is $2b_1(C_2)$ is proportional to both, the self-intersection $(C_2)^2$ and the tension of the D3-brane at the attractor value. 
As a consequence, the number of deformations, $b_1$, of the SLag grows to leading order quadratically in $N$ for large classes or when we rescale $C_2\rightarrow N C_2$:
$$b_1({\rm SLag2})\sim N^2 C_2^2.$$
\subsection{\texorpdfstring{Case II: SLag3 in CY$_3$}{Case II: SLag3 in CY3}}\label{ssec:n3}
Consider now type IIB string theory compactified on a Calabi--Yau three-fold $X_3$. This yields an effective theory in four dimensions with $\cN=2$ supersymmetry. Let $L_3\in H_3(X_3)$ be a special Lagrangian submanifold of $X_3$ and consider the BPS state obtained by wrapping a D3-brane on $L_3$. Since this state is BPS, its mass is given by the central charge 
\begin{align}\label{BHcentralcharge}
    \frac{M^2}{M_{P,4}^2}=8\pi |Z(L_3)|^2 =8\pi \left|\int_{L_3}\widehat{\Omega}_3\right|^2\,,
\end{align}
where $M_{P,4}$ is the four-dimensional Planck mass. Further consider the black hole solution associated to this BPS state which can be described via the attractor mechanism \cite{Ferrara:1995ih}. The attractor mechanism ensures that at the horizon of the 4d extremal black hole its central charge is minimized and for it to be a non-singular solution we need that the minimum value of $|Z(L_3)|\not= 0$.  Suppose this is indeed the case. We may then ask how $b_1(L_3)$ scales under rescaling of the class $L_3\rightarrow NL_3$.  Naively we may have thought that since the SLag is mid dimensional, just as in the case of $K3$ it should scale like the self-intersection of the cycle leading to an $N^2$ growth as before.  However, since the cycle is three-dimensional its self-intersection trivially vanishes. Nevertheless we now show that the expectation of an $N^2$ growth is still correct and that, again, there is an upper bound on the growth of $b_1$ proportional to $N^2$.  We do this by relating $b_1(L_3)$ to the spin of the BPS black hole.

For a 4d black hole of mass $M$, the spin is bounded by $8\pi J\leq M^2/M_{\rm{P},4}^2$ (see e.g. \cite{Wald:106274}). For a macroscopic BPS black hole in 4d, there is no spin.  However, this does not mean that the microscopic state does not carry a spin, but that the bulk of the microscopic states have no spin. Still, there can be spinning microstates with a lower growth than the bulk entropy. We will now use this to put a bound on $b_1$ by relating it to the spin of these microstates.

The D3 brane wrapped on SLag3 will in general carry a spin $J$ under the 4d $SO(3)\cong SU(2)$ rotation group. The possible spins $J$ are determined by the cohomology group of the moduli space $\widehat{\cM}$ of the D3-brane wrapped on $L_3$. The space $\widehat{M}$ is K\"ahler and of dimension $2 b_1(L_3)$ (as we can turn on Wilson lines) and therefore its cohomology allows for a Lefschetz decomposition where the $SL(2)$ acts on it for which the raising operator is given by the mutiplication with the K\"ahler class $J_{\widehat{\cM}}$ on $\widehat{\cM}$. The four-dimensional $SU(2)$ spin group can be identifed with the Lefschetz $SL(2)$. The spin $J$ then corresponds to the eigenvalue of the Cartan generator acting on $H^{p,q}(\widehat{M})$ which is given by 
\begin{align}
    (p+q-\text{dim}_{\mathbb{C}}\widehat{\cM})/2\,. 
\end{align}
As mentioned before, for the D3-brane on $L_3$ the moduli space $\widehat{\cM}$ is given by the supersymmetry-preserving deformations of $b_1(L_3)$ making it a complex d-dimensional space (after including the Wilson line). For a given $L_3$ the highest spin is thus
\begin{align}\label{Jmax}
    J_{\text{max}}=\frac{b_1(L_3)}{2}\,. 
\end{align}
For four-dimensional spinning black holes with fixed mass $M$, the spin further needs to satisfy the extremality bound 
\begin{align}
    8\pi J\leq \frac{M^2}{M_{{\rm P},4}^2}\,. 
\end{align}
Using that the mass of the D3-brane black hole is given by the value of the central charge at the attractor point $8\pi |Z(L_3)|\bigr|_{\rm min}$, we can apply the extremality bound to $J_{\rm max}$ in \eqref{Jmax} to find 
\begin{align}\label{b1L3}
    b_1(L_3) \leq \, 2|Z(L_3)|^2_{\rm min}\,.
\end{align}
 We thus see that the bound on $b_1$ can at most grow by a factor of $N^2$ as we rescale the class by a factor of $N$, as we had anticipated.
\subsection{\texorpdfstring{Cases IV, V$_1$: M5 on hol(C$_4$) in CY$_3$, CY$_4$}{Cases IV, V1: M5 on hol(C4) in CY3, CY4}}\label{ssec:hol4}

Let us now turn to the M-theory cases and consider M5-branes wrapped on holomorphic $4$-cycles of $CY_3$ and $CY_4$-folds. In that way we obtain a 2D supersymmetric theory with $\cN=(0,4)$ and $\cN=(0,2)$ supersymmetry, respectively.  In the IR these flow to SCFTs. Using the anomaly inflow we can now compute the $c_L,c_R$ of the resulting SCFT (using some genericity assumptions) which gives us an estimate of the degrees of freedom of the theory.  We compute this by noting that the level of the $U(1)$ R-symmetry (which in the first case is a Cartan of an $SU(2)$ R-symmetry) leads to $c_R$.  Moreover the gravitational anomaly leads to $c_L-c_R$.  We use the anomaly inflow on the M5 brane to compute these quantities.
In doing so we will assume that in the IR there are no accidentally enlarged R-symmetries.  This we expect to be the generic case when the 4-cycles are ample and movable inside the CYs (as would lead to cases with no enhanced symmetries) which is what we need in order to obtain AdS$_3$ solutions in the first place. 

To compute the anomaly inflow we need to integrate the anomaly polynomial $I_8$ for M5-branes over the internal 4-cycle $C_4$.  Let the M5-brane's world-volume $M$ be a 6-manifold which is locally a product of a very ample holomorphic 4-cycle, $C_4$, in $CY_4$ and a two-manifold.
The two-dimensional SCFT lives on the latter.
Let $T$ and $N$ denote the tangent bundle and the normal bundle to the M5 brane on $M$.  There are two possibilities: For $CY_3$ $N=P\oplus S$ where $P$ is the normal bundle in $CY_3$ and $S$ is the orthogonal complement $\C\oplus \mathbb{R}$, the supersymmetry is enhanced to $\cN=(0,4)$ and the $SO(3)$ R-symmetry is coupled to $S$. For the $CY_4$ case $N$ has a trivial one-dimensional part from $S$ which is not coupled to the R-symmetry (which we will neglect) and the rest (which is the normal bundle in $CY_4$) usually does not split and is coupled to the 2D R-symmetry line bundle $F$ given by the commutator (and center) $U(1)$ of $U(2)$ in $SO(5)$.  Note this R-symmetry is broken for a compact Calabi-Yau due to KK modes, but we still can use it to compute the inflow, as in the IR limit the KK modes are expected to decouple.

The anomaly 8-form for M5 branes is given by \cite{Freed:1998tg,Harvey:1998bx} \begin{align}
    I_8=\frac{1}{48}\left[p_2(N)-p_2(T)+\frac{1}{4}(p_1(N)-p_1(T))^2\right].
\end{align}
We perform the integral of this 8-form over $C_4$ separately for the two cases.
First when we turn off the 2D $R$-symmetry gauge field we may compute directly (using standard relations between Pontryagin classes) that
\begin{align}
    I_8=-\frac{1}{96}p_1(CY_4)p_1(T_2)=\frac{1}{48}c_2(CY_4)p_1(T_2) \,,
\end{align}
which implies that the two-dimensional SCFT satisfies $c_L-c_R=\frac{1}{2}\int_{C_4}c_2(CY_4)$ (corresponding to 24 times the coefficient of $p_1(T_2
)$). This holds for both $CY_3\times \C$ and strict $CY_4$ with $SU(4)$ holonomy.

To compute $c_R$ we need to turn on the R-current, which means that for strict $CY_4$ we replace $N$ by $N\otimes F$ and for the $CY_3$ case we replace $N$ by $P\oplus S$. By general properties of 2D $(0,2)$ SCFTs, in the first case the coefficient of the $c_1(F)^2$ term in the anomaly polynomial then gives $\frac{1}{6}c_R$ (a factor of $2$ comes from reading the level of $U(1)$ R-charge and a factor of 3 from the relation between the level of the $U(1)$ R-charge and the $c_R$). Similarly for $(0,4)$ theories we may fix the normalization by $c_1(F)^2$ term of the Cartan subgroup $U(1)\subset SO(3)$.

For $CY_3$, we find the $p_1(S)$ coefficient in $48I_8$ to be given by\footnote{  Keeping only the $p_1(S)$ term:\begin{align*}
    48(I_8-I_{8,p_1(S)=0})&=p_1(S)p_1(P)+\frac{1}{4}(p_1(CY_3)-2p_1(P)-p_1(S))^2\\
    &=p_1(S)(2C_4^2+c_2(CY_3))
\end{align*}}
$2C_4^2+c_2(CY_3)$, which (upon integration over $C_4$) gives  $c_R=(C_4)^3+\frac{1}{2} c_2(CY_3)\cdot C_4$.  This leads to 
\begin{align}
   c_R = (C_4)^3 + \frac12 c_2(CY_3) \cdot C_4 \,,\qquad  c_L = (C_4)^3 + c_2(CY_3) \cdot C_4 \,,
\end{align}
in agreement with \cite{Maldacena:1997de}.  Note that the fact that the leading behavior of $c_R,c_L$ upon rescaling $C\rightarrow NC$ goes as $N^3C^3$ can be naturally understood by deforming the M5 brane and interpreting the triple intersection of M5 branes as contributions to the degrees of freedom on the resulting string.  

For holomorphic very ample $C_4$ in $CY_4$ the coefficient of $c_1(F)^2$ in $48I_8$ is given by%
\footnote{Keeping only the $c_1(F)^2$ term:\begin{align*}
    48I_8&=(c_2(N)+c_1(N)c_1(F)+c_1(F)^2)^2+\frac{1}{4}(-2c_2(CY)+2c_2(N)-(c_1(N)+2c_1(F))^2\\
    &+2(c_2(N)+c_1(F)c_1(N)+c_1(F)^2)-(c_1(N)+2c_1(F))^2)^2\\
    &=c_1(F)^2(2c_2(N)+4c_2(T)-2c_2(CY))
\end{align*}
}
$2c_2(N)+4c_2(T)-2c_2(CY)$, which gives $c_R=\frac{1}{8}(2C_4\cdot C_4 +4\chi -2\int_{C_4} c_2(CY_4))$, leading to
\begin{align}\label{eq:Crclhol}
c_R=\frac{1}{4}\left(C_4\cdot C_4 +2\chi -\int_{C_4} c_2(CY_4) \right) \,,\quad c_L=\frac{1}{4}\left(C_4\cdot C_4 +2\chi +\int_{C_4} c_2(CY_4)\right) \,.
\end{align}

These formulas suggest that for large $C_4$ or upon rescaling $C_4\rightarrow NC_4$, the growth of $c_L,c_R$ scales as $N^2C_4^2/4$.  To see that the $\chi $ term does not spoil this note that if we deform $N$ copies of $C_4$ the only reason the central charges would not add is if they intersect. There are $N^2 C_4^2$ intersection points and each should contribute some universal number to $c_L,c_R$ consistent with this interpretation.

Thus, again, the growth of the number of degrees of freedom for mid-dimensional cycles scales in the leading order by a factor of $N^2$ upon rescaling the class by a factor of $N$.

\subsection{\texorpdfstring{Cases III,V$_2$: SLag4 in CY$_4$}{Cases III,V2: SLag4 in CY4}}\label{ssec:n4}
Let us now turn to the cases most relevant for the discussion in this paper, i.e. the cases III and V$_2$ which are deeply related. Case III involves studying D5/NS5 branes on orientifold of CY3, which, in turn, is more generally formulated as general $(p,q)$ 5-branes in the base of F-theory on an elliptic $CY_4$. Upon circle compactification, this then relates to M5 branes wrapping a SLag4, $L_4$, in $CY_4$.  Since the geometry is most easily described in the M-theory setup, we will use this geometric language and discuss what we expect in that case. 

To find the UV degrees of freedom of the worldvolume theory of the M5-brane on $L_4$, let us consider the reduction of the 6D M5-brane worldvolume theory on $L_4$ focusing on the bosonic sector only. The reduction of the chiral two-form in the 6D $(2,0)$ tensor multiplet yields $b_2^-$ left-moving and $b_2^+$ right-moving scalars. In addition, the tensor multiplet also contains five scalars, four of which describe the supersymmetry-preserving deformations of $L_4$ inside $X_4$ and the last one describing the motion of the resulting domain wall in the three extended directions. The tangent space of the deformations space $\mathcal{M}$ of $L_4$ is given by 
\begin{align}
    T_{L_4}(\cM) = H^0(L_4, \mathcal{N})= H^0(L_4, T^*L_4)\,,
\end{align}
where we used that $L_4$ being special Lagrangian implies the relation $\cN \cong T^*L_4$ for its normal bundle $\cN$. Using $H^0(L_4, T^*L_4)\simeq H^{1,0}(L_4)$, we find 
\begin{align}
    \text{dim}_\mathbb{R} \mathcal{M} = b_1(L_4)\,. 
\end{align}
The total number of left- and right-moving scalars is thus given by 
\begin{align}
    N_L= 1 + b_2^- + b_1 \,,\qquad N_R = 1+b_2^+ + b_1 \,,
\end{align} 
which pair with the fermions to form $N_L$ and $N_R$ $(1,1)$ multiplets leading to 
$(c^{UV}_L,c^{UV}_R)=\frac32(N_L,N_R)$.  Notice that, unlike in the case of holomorphic 4-cycles, we cannot determine $c_L,c_R$ individually by an anomaly inflow argument because there is no R-symmetry in the $(1,1)$ supersymmetric 2d theory. However, the difference $c_L-c_R$ can still be computed as before yielding  $\frac12 c_2(CY)\cdot L_4$.  For a Lagrangian $L_4$ it can be shown that this is given by $\frac32\sigma (L_4)$, where $\sigma(L_4)$ is the signature of $L_4$, in agreement with what we have found.\footnote{$T_{CY}=T_L\otimes \C$, hence $c_2(CY_4)=p_1(L_4)$ whose integral on $L_4$ gives $3\sigma$.}

To gain more insight to the number of degrees of freedom $N_L$ we delve a bit more deeply into the geometry of $L_4$.  As already noted, the local geometry of the Lagrangian $L_4$ inside the $CY_4$ geometry is given by $T^*L_4$.  This implies that
\begin{align}
L_4\cdot L_4 =\chi(L_4)\,,
\end{align}
where the Euler characteristic of $L_4$ is given by 
\begin{align}\label{EulercharacteristicL}
    \chi(L_4) = 2+ b_2^+ + b_2^- -2b_1\,.
\end{align} 
The total UV central charge of the theory on the M5-brane wrapping the special Lagrangian four-cycle, defined as $c_L^{UV}+c_R^{UV}$, is thus given by 
\begin{align}\label{finaleq}
 c_{\rm UV} = \frac{3}{2} \left(2+b_2^++b_2^- +2 b_1 \right ) =\frac{3}{2}(L_4\cdot L_4+4b_1 )\,,
\end{align}

We now want to find the scaling behaviour of $c_{\rm UV}$ for large classes or as we re-scale\footnote{Note that even though we write this as a non-primitive class, the expectation is that the leading behaviour of the brane is a smooth function of the class itself and it should not matter whether it is primitive or not.} $L_4\rightarrow NL_4$.  In this case just as we have seen in the previous examples we expect that as we deform $NL_4$ to $N$ separate copies, they intersect at $N^2L_4\cdot L_4$ points, each of which will lead to a universal contribution to $c_{\rm UV}$. Thus we learn that $c_{\rm UV}$ should scale by a factor proportional to $N^2$ at leading order for large $N$. In particular $b_1$ cannot be larger than a universal multiple of the self intersection, or, in other words, $b_1\leq a L_{4}\cdot L_4$.
Using the monotonicity of $c$ under RG flow,  we thus find the important relation that asymptotically in large class $L_4$ limit
$$c_{\rm IR}\leq c_{\rm UV}\leq \b L_4\cdot L_4$$
for some universal constant $\b \sim O(1)$.  Note that using the tadpole condition \eqref{eq:Mtheorytadpole} this is bounded by 
$$L_4\cdot L_4 \leq \frac{\chi(X_4)}{12} \,,$$
and therefore
$$c_{\rm IR}\leq \frac{\b}{12}\chi (X_4) \,.$$
Notice that this scaling argument similarly applies 
to a stack of $N$ individual branes on $L_4$ since also in this case the leading contribution to the number of degrees of freedom is counted by the self-intersection of $L_4$. 

Even though we derived this in the context of M-theory on $CY_4$, we expect the same to be true for the corresponding F-theory domain walls, as they lead to M-theory strings upon wrapping the circle.  In other words, if we denote the number of degrees of freedom of the F-theory domain wall by $c_F$ we expect
$$c_F\leq \gamma \chi(X_4) \,, $$
where $\gamma$ is some universal constant.  In principle one may have worried whether compactifying on the circle may lead to losing degrees of freedom (similar to M5 branes on a circle giving D4 branes). That this is not the case follows from the fact that $N$ coincident $(p,q)$ 5-branes has degrees of freedom also scaling as $N^2$, consistent with this bound. Moreover note that the argument of KKLT can be made directly in 3d (regardless of the 4d compactification) leading to the same issues.

To see this bound in explicit examples, we first consider the special case of $X_4=K3\times K3$ and then turn to the more general case of an arbitrary type IIB orientifold limit for which  $X_4 = (X_3 \times T^2)/\mathbb{Z}_2$ corresponding to the case III.

\subsubsection{\texorpdfstring{$K3\times K3$}{K3xK3}}\label{sec:K3xK3}
Take $X_4$  to be the product $K3\times K3$ of two $K3$ surfaces.
The special Lagrangian four-cycles of $X_4$ are now simply products of Riemann surfaces in each of the $K3$s. The SLag4-cycles are thus of the form
\begin{align*}
    L_4 =\Sigma^{g_1}\times  \Sigma^{g_2}\,,
\end{align*}
where $\Sigma^{g_1}$ is a genus-$g_1$ Riemann surface in $K3_1$ and $\Sigma^{g_2}$ a genus-$g_2$ Riemann surface in $K3_2$. 
The resulting two-dimensional theory has $\cN=(2,2)$ supersymmetry.
Since the Euler characteristic of a product manifold is the product of the Euler characteristics, we have 
\begin{align}
    \chi\left(\Sigma^{g_1}\times \Sigma^{g_2}\right) = (2g_1-2)(2g_2-2)\,. 
\end{align}
On the other hand from section \ref{ssec:n2} we recall that $b_1(\Sigma^{g_1})=2g_1$ such that 
\begin{align}
    b_1\left(\Sigma^{g_1}\times \Sigma^{g_2} \right) = 2g_1 +2g_2\,. 
\end{align}
Notice that the signature of $\Sigma^{g_1}\times \Sigma^{g_2}$ vanishes  
\begin{align}
    c_L-c_R = \frac{1}{2} \int_{\Sigma^{g_1}\times \Sigma^{g_2}} c_2(K3\times K3) =0 \,.
\end{align}
We therefore have $b_2^+=b_2^-=\frac{1}{2}b_2$ with 
\begin{align}
    b_2 = 4g_1 g_2 +2 \,,  
\end{align}
such that
\begin{align}\label{eq:cuvK3xK3}
    c_{UV} = \frac{3}{2} \left(4+4g_1 g_2 + 4g_1+4g_2\right)=6(g_1+1)(g_2+1) \,,
\end{align}
which for large $(g_1,g_2)$ scales like 
\begin{align}
    c_{UV} \sim 6 g_1 g_2 \,.
\end{align}
On the other hand, in this limit the self-intersection of $\Sigma^{g_1}\times \Sigma^{g_2}$ scales like 
\begin{align}
    \chi(\Sigma^{g_1}\times \Sigma^{g_2})\sim 4 g_1 g_2 \,.
\end{align}
Hence, $c_{UV}$ scales parameterically like the self-intersection in accordance with our general expectation for Calabi--Yau four-folds. 

Note that we can also view this example as holomorphic cycles, by suitable choice of complex structure on $K3$.  In that case the computations of the anomaly inflow \refeq{eq:Crclhol} leads to 
$c_{IR}=6(g_1-1)(g_2-1)$, which as expected is smaller than $c_{UV}$ by subleading terms in $g_i$.

\subsubsection{Orientifold limit}
Consider now the case III, i.e.\ take $X_4$ to be an orientifold $X_4 = (X_3 \times T^2)/\mathbb{Z}_2$, where $X_3$ is a CY 3-fold. In this case we can provide further evidence for the behaviour of $b_1$ using the black hole argument presented in section~\ref{ssec:n3}.

Let us consider a special Lagrangian sub-manifold $L_4$ of $X_4= (X_3 \times T^2)/\mathbb{Z}_2$ dual to a $G_4$ flux on $X_4$. In order for this flux to lift to a type IIB three-form flux \eqref{G4flux} tells us that $[L_4]$ needs to be of the form 
\begin{align}
    [L_4] = (C_R,A)+ (C_{NS},B) \,, 
\end{align}
with $(A,B)$ the one-cycles on $T^2$ and $C_{R}, C_{NS} \in H_3(X_3)$ are the three-cycles dual to the IIB three-form fluxes $F_3$ and $H_3$ and satisfy $C_R.C_{NS}>0$ .

In the following, we want to show two important properties of the classes $[L_4], C_R$ and $C_{NS}$. Firstly, we want to show that $L_4$ being a special Lagrangian implies that $C_R$ and $C_{NS}$ (or in fact multiples thereof) can also be represented by special Lagrangian three-cycles in $X_3$. Physically, this implies that, in the type IIB limit of M-theory, both the D5-brane and NS5-brane need to be wrapped in homology classes that have a special Lagrangian representative.  Secondly, we want to show that $b_1(L_4)$ is bounded by the dimension of the deformation space of $C_R$ and $C_{NS}$ as 
\begin{align}\label{boundb1L}
    b_1(L_4)\leq \text{min}\Bigl\{b_1(C_{R}), b_1(C_{NS})\Bigr\}+1\,.
\end{align}
This allows us to provide a bound on $b_1(L_4)$ through a bound on either $b_1(C_{NS})$ or $b_1(C_R)$.
\newline 

Let us start by showing that $C_R$ and $C_{NS}$ need to have special Lagrangian representatives if $[L_4]$ is special Lagrangian: take the $S^1$ family, $N_t$, of special Lagrangian cycles in $T^2$, such that $[N_t]=B$ and consider the intersection $L_R(t)= L_4\cap (X_3 \times N_t)$. This intersection has expected dimension three and satisfies the class condition $[L_R(t)]=C_R$. The three-cycles $L_R(t)$ are special Lagrangian themselves. To see this, let $p(t)$ denote a point on $T^2$ where $L_R(t)$ and $T^2$ intersect.  Let $v(t)$ denote the tangent vector of $L_R$ at $p$ in the $T^2$ direction.   We first note that 
\begin{align}
    \Omega_4\bigr|_{L_4} = \Omega_3\bigr|_{L_R(t)} \times \Omega_1\bigr|_{v(t)}\,.
\end{align}
Since by assumption $L_4$ is special Lagrangian and $\Omega_1\bigr|_{v(t)}$ does not depend on the points on $L_R(t)$ it follows that also $\Omega_3|_{L_R(t)}$ is constant. Similarly for the K\"ahler form $J_{X_4}$ of $X_4$ we have $J_{X_4} = J_{X_3} + J_{T^2}$ with $J_{X_3}$ the K\"ahler form of $X_3$ and $J_{T^2}$ the K\"ahler form of $T^2$. We further know that both, $J_{X_4}$ and $J_{T^2}$, restrict to zero along $L_R(t)$ such that also $J_{X_3}$ has to be zero along $L_R(t)$. Hence, we have an $S^1$ family of special Lagrangian three-cycles in  $X_3$ with $[L_R(t)]=C_R$. A second $S^1$ family of special Lagrangian three-cycles $L_{NS}(t)$ with $[L_{NS}(t)]=C_{NS}$ can then be constructed along the same lines.\footnote{In fact the above construction applies to $X_3\times T^2$. Taking into account the $\mathbb{Z}_2$ involution, we should consider the pre-image $[\tilde L] = (\tilde C_{R},A)+ (\tilde C_{NS},B)$ of $L$ in $X_3\times T^2$. Here $\tilde C_i = C_i -\sigma(C_i)$, with $\sigma$ the involution on $X_3$. Restricting to three-cycles $C_i$, that are anti-symmetric under the involution, we can apply the above construction to $[\tilde L] = (2C_{R},A)+ (2C_{NS},B)$ to find special Lagrangian representatives of the classes $2C_R$ and $2C_{NS}$.}

We thus showed that if $L_4$ is a special Lagrangian submanifold of $X_4$ the classes $C_R,C_{NS}\in H_3(X_3)$ also need to be represented by special Lagrangian submanifolds. Furthermore, the above construction shows that $L_4$ can be viewed as a fibration over $S^1$ in two different ways, once with fiber $L_R(t)$ and once with fiber $L_{NS}(t)$. The Serre spectral sequence\cite{serre1951homologie} thus tells us that $b_1(L_4)\leq b_1(L_R)+b_1(S^1)$ and $b_1(L_4)\leq b_1(L_{NS})+b_1(S^1)$. Defining $b_1(C_i)=\text{max}\left\{b_1(L_i)|[L_i]=C_i\right\}$ the relation \eqref{boundb1L} follows. 
\newline 

To find a bound on $b_1(L_4)$ we hence need to find an upper bound on either $b_1(C_{R})$ or $b_1(C_{NS})$. We now argue for such an upper bound based on the black hole argument of section~\ref{ssec:n3}. We showed above that $C_R$ and $C_{NS}$ (or more presicely $2C_R$ and $2C_{NS}$) need to have representatives, $L_R$ and $L_{NS}$, that are special Lagrangian submanifolds of $X_3$. As in section~\ref{ssec:n3} we can thus consider the black hole solutions associated to D3-branes wrapped on these submanifolds. Let us assume that the black hole exists for both, $L_R$ and $L_{NS}$,\footnote{Note that the existence of a black hole for $L_R$ and $L_{NS}$ is in fact not a necessary condition for $|Z(L_4)|$ to have a non-trivial extremum, as for example the conifold example we discussed shows.  However, the genericity assumption in the directions of NS and R fluxes which we are using throughout, as the statistical arguments of KKLT require it, show that this can be assumed.} and let us denote their respective attractor points in the complex structure moduli space of $X_3$ by $z^R,z^{NS}$. We will refer to the holomorphic three-form at these points in moduli space as $\Omega_3^R$ and $\Omega_3^{NS}$, respectively. Notice that in general $z^R\neq z^{NS}$. From the arguments presented in section~\ref{ssec:n3} it follows that 
\begin{align}\label{boundorienti}
    b_1(L_{R}) \leq 2 \, \bigl|Z(L_R)\bigr|_{z^R}^2 \,, \qquad  b_1(L_{NS}) \leq 2\, \bigl|Z(L_{NS})\bigr|^2_{z^{NS}} \,. 
\end{align}
Thus, the existence of a BPS black hole solution for D3-branes on $(L_{R},L_{NS})$ provides us with a bound on $b_1(L_{R, NS})$. 

However, this is not yet what we are looking for since we aim to bound $b_1(L_4)$ in terms of its self-intersection $\chi(L_4)$.
Let us denote the point in moduli space at which $|Z(L_4)|$ is extremised by $(\tau^*, z^*)$ corresponding to the value of the type IIB axio-dilaton and a point in the complex structure moduli space of $X_3$ at which the holomorphich $(3,0)$ form is given by $\Omega_3^*$. Notice that in general $z^*$ differs from both $z^R$ and $z^{NS}$. However, since $|Z(L_R)|$ is minimized by $\Omega_3^R$ and $|Z(L_{NS})|$ by $\Omega_3^{NS}$, we have
\begin{eqn}\label{boundc1Rone}
     \left|\int_{L_R} \widehat{\Omega}_3^R\right|^2 &\leq \left|\int_{L_R} \widehat{\Omega}_3^*\right|^2 \,,\qquad
      \left|\int_{L_{NS}}\widehat{\Omega}_3^{NS}\right|^2&\leq\left|\int_{L_{NS}} \widehat{\Omega}_3^*\right|^2
\end{eqn} 
In terms of the dual fluxes $F_3$ and $H_3$ the F-term conditions, i.e.\ the extremisation condition for $|Z(L_4)|$, imply 
\begin{align}
    0 = \int_{X_3} G_3\wedge\bar \Omega_3^* = \int_{X_3} F_3 \wedge \bar\Omega_3^* - \tau^*\int_{X_3} H_3\wedge \bar \Omega^*_3\,,
\end{align}
where we used that at the attractor point of $L_4$, the flux $G_3$ can only have $(2,1)$ and $(0,3)$ components. Together with \eqref{boundc1Rone} this relation yields
\begin{eqn}\label{boundCRCNS}
    b_1(L_R) &\leq 2\left|\int_{X_3} F_3 \wedge\widehat{\Omega}_3^*\right|^2 =2\left|\frac{\tau^*}{2\tau_2^*}\right|^2 \left|\int_{X_3}  G_3 \wedge  \widehat{\Omega}_3^*\right|^2\,,\\
    b_1(L_{NS}) &\leq 2\left|\int_{X_3} H_3 \wedge \widehat{\Omega}_3^*\right|^2 =2\left(\frac{1}{2\tau_2^*}\right)^2 \left|\int_{X_3}  G_3 \wedge  \widehat{\Omega}_3^*\right|^2\,,
\end{eqn}
where we split $\tau$ into its real and imaginary part, $\tau = \tau_1 + i \tau_2$.
Now, we can use 
\begin{align}
    \int_{X_3} G_3\wedge \Omega_3 = \int_{X_4 }G_4 \wedge \Omega_4 \,,\qquad 2i \tau_2 \int_{X_3} \Omega_3\wedge \bar{\Omega}_3 = \int_{X_4} \Omega_4\wedge\bar\Omega_4 \,,
\end{align}
to rewrite \eqref{boundCRCNS} as
\begin{align}
    b_1(L_R) &\leq  \frac{\left|\tau^*\right|^2}{\tau_2^*} \left|\int_{X_4} G_4 \wedge\widehat{\Omega}_4^*\right|^2= \frac{\left|\tau^*\right|^2}{\tau_2^*}\bigl|Z(L_4)\bigr|^2_{\rm min} \,,\\
    b_1(L_{NS}) &\leq \frac{1}{\tau_2^*}\left|\int_{X_4} G_4 \wedge \widehat{\Omega}_4^*\right|^2=\frac{1}{\tau_2^*}\bigl|Z(L_4)\bigr|^2_{\rm min}\,. 
\end{align}
Using \eqref{boundb1L} we can finally translate these bounds into a bound on $b_1(L_4)$. Therefore, we need to identify the $\text{min}\left\{(b_1(L_{R}),b_1(L_{NS})\right\}$. To that end notice that $\tau^*$ takes values in the fundamental domain of $SL(2,\mathbb{Z})$ such that
\begin{align}\label{boundontau}
    \tau_2^* \geq \frac{\sqrt{3}}{2}\,. 
\end{align}
 On the other hand, \eqref{boundontau} allows us to bound $b_1(L_{NS})$ as
\begin{align}
    b_1(L_{NS}) \leq \frac{2}{\sqrt{3}}\left|Z(L_4)\right|^2_{\rm min}\,. 
\end{align}
By \eqref{boundb1L} this is sufficient to give an upper bound on $b_1(L)$ as
\begin{align}
    b_1(L_4) \leq \frac{2}{\sqrt{3}} \left|Z(L_4)\right|^2_{\rm min}+1 \,, 
\end{align} 
which can again be rewritten as a bound on $b_1(L_4)$ in terms of self-intersection $\chi(L_4)=L_4.L_4$ by noticing that $|Z(L_4)|^2\leq \chi(L_4)$ at the attractor point where the flux dual to $L_4$ is self-dual, i.e.
\begin{align}\label{b1boundorienti}
    b_1(L_4) \leq \frac{2}{\sqrt{3}} \chi(L_4)+1 \,. 
\end{align} 

To summarize, using \eqref{finaleq}, we have found a bound on the number of degrees of freedoms of M5-branes on SLag four-cycles $L_4$ in terms of the self-intersection of $L_4$ . In particular, our analysis shows that under the re-scaling $L_4 \rightarrow N L_4$, the degrees of freedom scale as 
\begin{align}\label{Cuvgrowth}
    c_{\rm UV}(NL_4) \sim N^2 c_{\rm UV}(L_4)\,. 
\end{align}

A further consistency check on our results can be seen as follows: consider M-theory on a general CY four-fold $\tilde{X}_4$ with an M5-brane wrapping a holomorphic four-cycle $C_4$ as in section \ref{ssec:hol4}. If we reduce the effective three-dimensional theory on a further $S^1$ we arrive at type IIA  on $\tilde{X}_4$. If the M5-brane wraps the additional $S^1$ it gets mapped to a D4-brane in type IIA on $C_4$. Mirror symmetry for type IIA in two dimensions maps the D4-brane on $C_4$ to a D4-brane on a special Lagrangian submanifold $L_4$ of the mirror four-fold $X_4$. Lifting type IIA on $X_4$ to M-theory on $X_4$, the D4-brane on $L_4$ gets lifted to an M5-brane on $L_4$. The degrees of freedom on the M5-brane on $C_4$ can now be identified with the degrees of freedom of the M5-brane on $L_4$ such that from \eqref{eq:Crclhol} we learn that, upon rescaling $L_4\rightarrow NL_4$, the central charge grows as \eqref{Cuvgrowth}. Thus, again, we see that the degrees of freedom on the four-cycles mirror to holomorphic four-cycles essentially grows as the self-intersection of the special Lagrangian. 

\section{ Holographic obstruction for the KKLT scenario} 
We now return to the holographic description of supersymmetric AdS vacua in type IIB/M-theory flux compactifications. In particular, we wish to relate the behavior of the degrees of freedom on the M5-brane domain walls to the possibility of finding supersymmetric AdS vacua with exponentially small cosmological constant as required for the KKLT scenario reviewed in section \ref{sec:KKLT}. 

Recall from section \ref{sec:holography} that the degrees of freedom on the domain wall give an upper bound on the degrees of freedom of the CFT dual to the supersymmetric AdS vacuum. In $d$ dimensions the number $c$ of degrees of freedom of the CFT are related to the AdS radius via 
\begin{align}
    c \sim l_{\rm AdS}^{d-2}\,. 
\end{align}
Thus, in order for the AdS radius to be very large (i.e. small absolute value for the cosmological constant) we need $c$ to be very large. In order to realize the KKLT scenario we thus have to search for a 5-brane configuration with very large $c_{\rm UV}$. For definiteness, let us focus first on the three-dimensional case, i.e.\ $CY_4$-fold compactifications of M-theory. For this case we showed in section \ref{sec:SLag} that the UV degrees of freedom of an M5-brane dual to a supersymmetric AdS flux vacuum are bounded by the self-intersection of the special Lagrangian four-cycle $L_4$ wrapped by the M5-brane, i.e. 
\begin{align}
    c_{\rm UV} \leq \beta \chi(L_4)\,. 
\end{align}
Within the tadpole bound, we can in principle maximize this by considering $L_4$ such that 
\begin{align}
    \frac12 \chi(L_4) = \frac{\chi(X_4)}{24}\,.
\end{align}
Physically this means that we consider a flux that exactly cancels the tadpole induced by the curvature of the CY$_4$-fold without having to introduce additional M2-branes. For instance for the $K3\times K3$ example discussed in section \ref{sec:K3xK3} one could choose $L_4 = \Sigma^{g_1} \times \Sigma^{g_2} $ with $48 = (2g_1 -2) (2g_2-2)$, i.e.~$(g_1,g_2)\in \left\{(13,2),(7,3),(5,4),(4,5),(3,7),(2,13)\right\}$. Then from \eqref{eq:cuvK3xK3} we have 
\begin{align}
    c_{\rm UV} \leq 252 \,. 
\end{align}
Thus, we only expect supersymmetric AdS$_3$ vacua from flux compactifications on $K3\times K3$ with AdS radius
\begin{align}
    l_{\text{AdS}_3}(K3\times K3) \leq c_{\rm UV}=252\,. 
\end{align}
If we consider more general elliptic CY four-folds with large Euler characteristic we might then hope to find vacua with even smaller cosmological constant. The CY four-fold with largest Euler characteristic known currently has been constructed in \cite{Klemm:1996ts} and was further discussed in detail in \cite{Taylor:2015xtz}. For this particular manifold we have
\begin{align}
    \chi(X_4^\text{max}) = 1\,820\,448\,, 
\end{align}
which provides an upper bound for the Euler number of Fermat-type CY fourfolds \cite{Klemm:1996ts}. According to a statistical argument in \cite{Taylor:2015xtz} this CY allows for a vast landscape of flux vacua. Based on our discussion in section \ref{sec:SLag} the degrees of freedom of an M5-brane wrapped on a special Lagrangian submanifolds of $X_4^\text{max}$ can be as large as 
\begin{align}
    c_{\rm UV} \lesssim \beta \frac{\chi(X_4^{\rm max})}{24} \sim \mathcal{O}(10^5)\,,
\end{align}
where $\beta\sim \mathcal{O}(1)$. For this example we thus expect supersymmetric AdS$_3$ vacua of the order 
\begin{align}\label{lAds3max}
    l_{{\rm AdS}_3}(X_4^\text{max}) \lesssim \mathcal{O}(10^{5})\,. 
\end{align}

We can repeat the above analysis for the four-dimensional AdS$_4$ vacua obtained from flux compactifications of type IIB string theory on Calabi--Yau orientifolds. In this case, the tadpole cancellation condition reads
\begin{equation}\label{eq:IIBtadpole}
    \frac12 \int F_3 \wedge H_3 +N_{D3} = \frac14 N_{O3} + \frac{1}{24} \bigl[\chi(D7) + 2 \chi(O7) \bigr] \,,
\end{equation}
where $N_{D3/O3}$ denotes the number of D3/O3 planes and $\chi(D7)$ and $\chi(O7)$ are the Euler characteristics of the divisors wrapped by D7 branes and O7 planes, respectively. The cancellation of the D7 tadpole requires that
$\chi(D7) = 4 \chi(O7)$ and \eqref{IIBtadpole} reduces to
\begin{equation}\label{eq:IIBtadpoleorienti}
    \frac12 \int F_3 \wedge H_3 +N_{D3} =  \frac14 \bigl[ N_{O3}+ \chi(O7) \bigr] \,.
\end{equation}
The rhs of the above equation corresponds to the Euler characteristic of the fixed point set of the orientifold projection which is related to $\chi\left(X_4\right)=\chi\left(X_3\times T^2)/\mathbb{Z}_2\right)$ in the following way. For a general CY manifold $M$ on which we act by an orbifold group $G$, the orbifold formula gives \cite{Dixon:1985jw}
\begin{align}
    \chi(M/G) = \frac{1}{|G|} \sum_{ g_1g_2=g_2g_1}\chi(g_1,g_2) \,,\qquad g_1,g_2\in G\,, 
\end{align}
where $\chi(g_1, g_2)$ corresponds to the Euler characteristic of the fixed point set under $g_1g_2$. For $M=X_3\times T^2$ and $G=\mathbb{Z}_2$ the orientifold action, the contribution from $g_1=g_2=1$ vanishes whereas from the other sectors we obtain $1/2$ times $3$ times the Euler characteristic of the fixed point set of $\mathbb{Z}_2$, i.e.  
\begin{align}\label{eq:orientifoldtadpole}
   \chi\left(\frac{X_3\times T^2}{\mathbb{Z}_2}\right)= \frac32 \,\chi\left(X_3|_{\rm f.p.}\times T^2|_{\rm f.p.}\right)\,=
   \,6 \chi\left(X_3|_{\rm f.p.}\right)\,,
\end{align}
where we used that the $\mathbb{Z}_2$ action has 4 fixed points on $T^2$.
This leads to 
\begin{align}\label{eq:eulerOrientifold}
   \frac{1}{24} \chi\left(\frac{X_3\times T^2}{\mathbb{Z}_2}\right)=
   \,\frac14 \chi\left(X_3|_{\rm f.p.}\right)\,=  \frac14 \bigl[ N_{O3}+ \chi(O7) \bigr] \,,
\end{align}
as expected.
We are thus left to calculate the Euler characteristic of the fixed-point set on $X_3$. By the Lefshetz fixed-point theorem this last contribution is given by (cf.\ also \cite{Carta:2020ohw,Bena:2020xrh})
\begin{align}
    \chi\left(X_3|_{\rm f.p.}\right) = \sum_i (-1)^i (b_i^+ - b_i^-) \,, 
\end{align}
where $b_i^\pm$ count the cohomology classes that are even or odd under the $\mathbb{Z}_2$ action. We thus get 
\begin{align}\label{chiorienti}
    \chi\left(X_3|_{\rm f.p.}\right) = 2\left(2 + (h^{1,1}_+ - h^{1,1}_-) - (h^{2,1}_+-h^{2,1}_-\-)\right)< 4 + 2\left(h^{1,1}+h^{2,1}\right)\,,
\end{align}
and therefore
\begin{align}
     \frac{1}{24} \chi\left(\frac{X_3\times T^2}{\mathbb{Z}_2}\right)< 1+\frac12 \left(h^{1,1}+h^{2,1}\right) < 252 \,,
\end{align}
where in the last step we estimated the maximal Hodge numbers based on the Kreuzer-Skarke list \cite{Kreuzer:2000xy}. We thus find a bound for the rhs of \eqref{eq:IIBtadpoleorienti}. 

From the analysis of section \ref{ssec:n4} we know that the degrees of freedom on the domain wall obtained from D5-/NS5-branes is bounded by its self-intersection number. Therefore the cosmological constant of the dual AdS$_4$ is bounded by the available tadpole which, given the previous discussion, is bounded by $\int F_3 \wedge H_3 \lesssim \mathcal{O}(500)$. We thus get the bound 
\begin{align}\label{lAds4max}
    l_{\text{AdS}_4} M_{{\rm P},4} \lesssim \sqrt{c_{UV}} \lesssim \mathcal{O}(22) \,.
\end{align}
We thus expect only AdS$_4$ vacua with cosmological constant at most $\Lambda/M_{P,4}^2 \sim \mathcal{O}(10^{-2})$ such that for type IIB orientifold we do not expect any supersymmetric AdS vacua with exponentially suppressed cosmological constant as envisioned by the KKLT scenario. Even if we consider the more general case of an elliptic fourfold, which would not lead to weak coupling control, we find 
\begin{align}\label{lAds4elliptic}
    l_{\text{AdS}_4} M_{{\rm P},4} \lesssim \sqrt{c_{\rm UV}}\sim \sqrt{10^6} \lesssim \mathcal{O}(10^3) \,.
\end{align}
We now wish to give an argument why even the supersymmetric AdS$_3$ and AdS$_4$ vacua with cosmological constants satisfying the bounds \eqref{lAds3max} and \eqref{lAds4max} obtained via duality cannot be under perturbative control. To this end note that the number of light degrees of freedom, $N_\text{light}$, below the string and KK-scale are given by the light multiplets associated to the moduli of the compactifications counted by $h^{3,1}$ and $h^{1,1}$ for the CY four-fold case and $h^{2,1}$ and $h^{1,1}$ in the orientifold case.\footnote{Notice that this is still true if these fields pick up mass due to the non-trivial scalar potential, since for consistency we expect the masses for complex structure and K\"ahler fields to be below the string scale (see e.g.~\cite{Kachru:2018aqn}).} For a CY four-fold $X_4$ we have 
\begin{align}
    \chi(X_4) = 6(8+h^{3,1} +h^{1,1} - h^{2,1})\,,
\end{align}
whereas for the orientifold case the relation between $\chi$ and the Hodge numbers can be estimated as in \eqref{chiorienti}. These are the minimum number of light states. We therefore in general have
\begin{align}
  N_{\rm light} \gtrsim \chi(X_4)\,,
\end{align}
up to $\mathcal{O}(1)$ coefficients. On the other hand, the species length scale $l_{\rm sp}$ in $d$-dimensions (for both $d=3,4$ cases of interest here) is given by 
\begin{align}\label{speciesscale} 
l_{\rm sp}^{d-2} = l_{ {\rm P},d}^{d-2}{N_{\rm light}} \gtrsim \chi(X_4)\ l_{ {\rm P},d}^{d-2}\,,
\end{align}
where $L_{{\rm P},d}$ is the Planck length in $d$ dimensions. Using \eqref{lAds3max} and \eqref{lAds4max} we then find 
\begin{align}
   l_{\text{AdS}_d}^{d-2}\lesssim \chi(X_4)\  l_{ {\rm P},d}^{d-2}\lesssim l_\text{sp}^{d-2} \qquad\Rightarrow\qquad \frac{l_{\text{AdS}_d}}{l_{sp}} \lesssim 1 \,, 
\end{align}
which is valid for both $d=3,4$. Hence, even in setups where it seems possible to get relatively small cosmological constants we do not get actual AdS$_3$ or AdS$_4$ vacua since the AdS length scale is always parameterically at or below the species length scale and thus the EFT breaks down. We thus do not expect to find any KKLT-like AdS$_3$ or AdS$_4$ vacua with exponentially small cosmological constant in a controlled regime of the EFT. Recall that here all our analysis is under the assumption that $|\mathcal{Z}|$ does have a non-trivial extremum. As already discussed in section \ref{sec:SLag} such an extremum can never be a global minimum since $|\mathcal{Z}|\rightarrow 0$ for $\mathcal{V}\rightarrow \infty$.\footnote{For the case of M5 branes wrapping holomorphic 4-cycles, it is clear that CY$_4$ will have to partially decompactify in the holographic dual, as the $U(1)$ R-symmetry of 2d SCFT with $(0,2)$ supersymmetry, demands a circle symmetry which would be absent for compact CY.} We thus have to assume that there are at least two different attractor points with different values for $|\mathcal{Z}|$. While one of the attractor points signals a decompactification of the theory, via holography and the species scale, we showed that any other attractor points (if they exist) are necessarily at strong coupling. 

\subsubsection*{Comparison to previous results}
In this section, we want to relate our results obtained via holography to previous attempts to find KKLT-like supersymmetric AdS vacua from full string theory constructions.
The main focus of research thus far has been on finding a small value of the superpotential by ignoring the superpotential dependence on K\"ahler moduli and focusing only on solving the complex structure equations, and hoping that the K\"ahler moduli corrections will not significantly alter the result of the minimum achieved by complex moduli stabilization.  For a completion of the first step one does need suitable K\"ahler moduli-dependent corrections to the superpotential as otherwise the overall volume factor in the $W$ will lead to supersymmetry being realized through decompactification. However the difficult task to compute the complete superpotential and K\"ahler potential including the K\"ahler moduli dependence has not been achieved.\footnote{For partial progress in this direction see \cite{Demirtas:2021nlu,Demirtas:2021ote}.}

A recent attempt in trying to get a small value of $W$ after complex structure variation was initiated by the study of the so-called perturbatively flat (flux) vacua first proposed in \cite{Demirtas:2019sip} and subsequently studied in \cite{Demirtas:2020ffz,Blumenhagen:2020ire,Demirtas:2021nlu,Demirtas:2021ote}. These attempts follow the original KKLT description to find supersymmetric AdS$_4$ vacua in type IIB string theory and first try to engineer flux compactifications which stabilize all complex structure moduli and the dilaton with small $W_0\equiv W|_{D_iW=0}$. To achieve a small $W_0$,  \cite{Demirtas:2019sip} proposes to first only consider the leading complex structure dependence of $\int_{X_3} \widehat{\Omega}\wedge G_3$ and search for fluxes, $G_3$, that are imaginary self-dual along a complex one-dimensional subspace of the moduli space with $\int_{X_3} \widehat{\Omega}\wedge G_3\bigr|_{*G_3=iG_3}=0$. At this level, there is thus a flat direction in moduli space, hence the name perturbatively flat vacua. If the exponential corrections to $\int_{X_3} \widehat{\Omega}\wedge G_3$, dual to non-perturbative worldsheet instanton corrections of the mirror, are taken into account, this flat direction is lifted and $G_3$ is imaginary self-dual only at isolated points in moduli space. As shown in \cite{Demirtas:2019sip} it is then possible that these corrections lead to
\begin{align}
0\neq \left.\left(\int_{X_3} \widehat{\Omega}\wedge G_3\right)\right|_{*G_3=iG_3}\ll 1\,. 
\end{align}
Starting from these flux configurations with exponentially small $W_0$, one then aims to realize the KKLT scenario and find supersymmetric AdS vacua when taking into account non-perturbative corrections to the superpotential from e.g.\ D3-brane instantons depending on K\"ahler moduli (cf.~\cite{Demirtas:2021nlu,Demirtas:2021ote}). We now want to argue that the perturbatively flat vacua of the type proposed in \cite{Demirtas:2019sip} cannot yield supersymmetric AdS vacua, consistent with our expectation from holography. As we will show the main problem with the construction of \cite{Demirtas:2019sip} is that fluxes considered there cannot be dualized into 5-branes wrapping special Lagrangian three-cycles. Based on our discussion in section \ref{ssec:n4} the corresponding four-cycle in $X_4=(X_3\times T^2)/\mathbb{Z}_2$ can also not be a special Lagrangian four-cycle. Therefore there is no $\frac12$-BPS domain wall associated to this choice of fluxes and supersymmetry is broken.  In other words suitable corrections to the superpotential depending in particular on K\"ahler moduli, which is hoped to lead to a supersymmetric vacuum, will not materialize.

To see that the cycles dual to the fluxes considered by \cite{Demirtas:2019sip} in general do no admit special Lagrangian representatives, let us briefly review  the setup of \cite{Demirtas:2019sip}. For a compactification of type IIB string theory on a Calabi--Yau orientifold $X_3$, \cite{Demirtas:2019sip} give a sufficient condition for the existence of a perturbatively flat direction which requires the superpotential $W$ to be a polynomial of degree-2 in the $h^{2,1}+1$ moduli describing the complex structure and the axio-dilaton. The flat direction then corresponds to the overall rescaling modulus. Let us denote the projective coordinates on the complex structure moduli space by $Z^I=(1,z^i)$, $I= 0, \dots, h^{2,1}$ such that 
\begin{align}
    \int_{A_i} \Omega_3 = z^i \,,\qquad \int_{B^i} \Omega_3 = F_i \equiv \partial_{z^i} \mathcal{F}\,,
\end{align}
with $(A_I,B^I)$ a symplectic basis of three-cycles on $X_3$. Moreover, $\mathcal{F}$ is the prepotential. For the moment consider only the contributions to $\mathcal{F}$ that are polynomial in the $z^i$, i.e. 
\begin{align}
    \mathcal{F}_{\rm pert} = -\frac{1}{3!}\kappa_{ijk} z^i z^jz^k +\frac{1}{2} a_{ij}z^iz^j +b_i z^i +\zeta\,,
\end{align}
where $\kappa_{ijk}$ are the triple intersection numbers of the mirror of $X_3$, $a_{ij}$ and $b_i$ are rational and $\zeta=-\frac{\zeta(3)\chi}{2(2\pi i)^3}$ and $\chi$ the Euler characteristic.  To get a superpotential that is homogeneous in the moduli, the fluxes $(F_3,H_3)$ need to be chosen such that 
\begin{align}
    W_\text{pert.} = -\tau K_i z^i - \frac{1}{2} \kappa_{ijk} z^iz^j M^k\,, 
\end{align}
which is achieved for the flux choice 
\begin{eqn}
    \left(\int_{A_I} H_3, \, \int_{B^I} H_3\right) &= \left(0, \vec{K}^T\,,\;0\,,\;0\right)\,,\\
     \left(\int_{A_I} F_3, \, \int_{B^I} F_3\right) &= \left(\vec{M}\cdot \vec{b}\,, \;\vec{M}^T\cdot \mathbf{a}\,, \;0\,,\;\vec{M}^T\right)\,. 
\end{eqn}
In this case, the D3-brane tadpole is simply given by 
\begin{align}
    Q_{D3}=-\frac{1}{2} \vec{M}\cdot \vec{K}\,.
\end{align}
According to \cite{Demirtas:2019sip} a perturbatively flat vacuum with $W_\text{pert}=0$ is obtained if $N_{ij}\equiv \kappa_{ijk} M^k$ is invertible and $\vec K^T \cdot \mathbf{N}^{-1} \cdot \vec{K} = 0$ provided $\vec{p}\equiv \mathbf{N}^{-1} \vec{K} $ lies in the K\"ahler cone of the mirror of $X_3$. The flat direction is then given by $\vec{z} = \tau \vec{p}$ with $\tau$ the axio-dilaton. In the scenario of \cite{Demirtas:2019sip} this remaining flat direction can then be stabilized using exponential corrections to the prepotential $\mathcal{F}$ dual to worldsheet instantons on the mirror of $X_3$. 

Even though by this choice of fluxes the F-term equations $D_i W = D_\tau W=0$ can be solved with $W_0\ll 1$, this scenario cannot yield a supersymmetric AdS vacuum since the three-cycles dual to the $(F_3,H_3)$ flux do not have special Lagrangian representatives at the point in moduli space where the F-term equations are satisfied. To see this, notice that, by construction, the cycle $C_{NS}$ dual to the $H_3$ flux satisfies along the flat direction
\begin{align}
    \int_{C_{\rm NS}} \Omega = \vec{K} \cdot \vec{z} = \tau \vec{K} \cdot \vec{p}= \tau \vec{K} \cdot \mathbf{N}^{-1} \vec{K} =0\,. 
\end{align}
Notice, that this equation is independent of exponential corrections to the prepotential because the $\vec{z}$ are the flat coordinates on the moduli space.
Therefore, it holds not only exactly at the large complex structure point $z^i \rightarrow \infty$ but also has a one-parameter family of solutions even if the corrections are relevant and taken into account.
Consequently, if $C_{\rm NS}$ had a special Lagrangian representative, type IIB compactified on $X_3$ would have a BPS state, obtained by wrapping a D3-brane on $C_{\rm NS}$ whose central charge would vanish along the flat direction. A point in moduli space at which a BPS state becomes massless, however, has to correspond to a singularity of the moduli space. Since by assumption $\vec{z}$ lies in the K\"ahler cone of the mirror of $X_3$, the flat direction corresponds to a smooth locus in moduli space. Therefore, in the vicinity of this locus, the D3-brane on $C_{\rm NS}$ cannot be BPS and the curve cannot be a special Lagrangian. A similar conclusion holds for the curve $C_{\rm RR}$. 

In fact, we can be more explicit. The condition $\vec{K} \cdot \vec{z}=0$ implies that some of the entries of $\vec{K}$ have to be negative since $\Im z^i>0$ inside the K\"ahler cone of the mirror of $X_3$. By construction the cycle wrapped by the NS5-branes dual to the $H$-flux is a three-cycle that via mirror symmetry gets mapped to a two-cycle $C$ in the class determined by $\vec{K}$. However, there are no holomorphic two-cycles in a class that is a linear combination of two-cycle classes with coefficients of non-definite sign. The relevant curves for the example studied in \cite{Demirtas:2019sip} can be described in terms of the generators $C_1$ and $C_2$ of the Mori cone of the Calabi--Yau threefold $\mathbb{P}_{1,1,1,6,9}[18]$. To be precise, the flux-choice of \cite{Demirtas:2019sip} corresponds to the class 
\begin{align}
    [C] = 3[C_1] - 4[C_2]\,. 
\end{align}
This class, however, does not have a holomorphic representative and therefore the dual three-cycle in the mirror of $X_3$ does not have a SLag representative. Therefore, the system of NS5-/D5-branes corresponding to the fluxes considered in \cite{Demirtas:2019sip} does not preserve supersymmetry and hence cannot be dual to a supersymmetric AdS vacuum. 
\section{Conclusions}

In this paper, using holography, we have argued, why the first step of KKLT scenario cannot be realized.
In a sense what we found is not that surprising and perhaps should have been expected: We have a dual theory involving of order of $\sqrt \chi$ D5/NS5 branes and thus to get a central charge bound of order $\chi$ is natural leading to an AdS length scale below or of the order of the species length scale.  In the usual AdS/CFT we do not have such a bound and we can have an arbitrarily large number of branes and an arbitrarily small negative cosmological constant, perhaps similar to a DGKT type scenario \cite{DeWolfe:2005uu}.  However in such a flux compactatification which has no such bounds on the cosmological constant, we expect to have a tower of light states \cite{Lust:2019zwm} and in this case the light tower of states prevents an uplift \cite{Kallosh:2006fm,Hertzberg:2007wc}.

Given the difficulties of realizing other scenarios for constructing dS solutions
in the context of flux vacua (see e.g.\cite{Junghans:2022exo} for some issues in the LVS context) it is natural to broaden the search for constructing quasi-dS vacua in string theory.  Whether or not we can construct meta-stable dS remains an open question in string theory. But regardless of that, a quasi-dS (not necessarily meta stable) should presumably be realizable in the string landscape given the observations in our universe.  This is one of the most urgent problems in string theory, and we hope the current work suggests looking with a fresh eye in different directions to realize this goal.

\subsubsection*{Acknowledgements}
We have greatly benefited from discussions with Mirjam Cvetic,  Mariana Gra\~na, Daniel Jafferis,  Hee-Cheol Kim, Miguel Montero, Gary Shiu, Irene Valenzuela, Thomas Van Riet and Timo Weigand.

The work of SL is supported by the NSF grant PHY-1915071.
The work of CV and MW is supported in part by a
grant from the Simons Foundation (602883, CV) and also by the NSF grant PHY-2013858.

\bibliography{papers_Max}
\bibliographystyle{JHEP}

\end{document}